\documentclass[onecolumn]{aastex6}

\newcommand{\beq}{\begin{equation}}
\newcommand{\eeq}{\end{equation}}

\usepackage{amsmath}
\usepackage{natbib}
\usepackage{amsfonts,amssymb}
\usepackage{color}
\usepackage{graphicx}

\begin{document}


\title{A comprehensive study on the variation phenomena of AO 0235+164}

\shorttitle{Variability mechanism }
\shortauthors{Wang \& Jiang}
%% Use \author, \affil, and the \and command to format author and affiliation
%% information.  If done correctly the peer review system will be able to
%% automatically put the author and affiliation information from the manuscript
%% and save the corresponding author the trouble of entering it by hand.
%%
%% The \affil should be used to document primary affiliations and the
%% \altaffil should be used for secondary affiliations, titles, or email.

%% Authors with the same affiliation can be grouped in a single
\author{Yi-Fan Wang\altaffilmark{1}, Yun-Guo Jiang\altaffilmark{1}}
\affil{Shandong Provincial Key Laboratory of Optical Astronomy and Solar-Terrestrial Environment,\\
Institute of Space Sciences, Shandong University, Weihai, 264209, China; jiangyg@sdu.edu.cn}

\begin{abstract}
The variation mechanism of blazars is a long standing open question. The polarization observation can provide us with more information to constrain models. In this work, we collect the long term multi-wavelength data of AO 0235+164, and make correlation analysis between them by using the local cross-correlation function (LCCF).  We found that both $\gamma$-ray and the optical {\it V}-band light curves are correlated with the radio light curve at beyond 3$\sigma$ significance level. The emitting regions of the $\gamma$-ray and the optical coincide within errors, and locate at  {$6.6_{-1.7}^{+0.6}$} pc upstream of the core region of 15 GHz, which are beyond the broad line region (BLR).
The color index shows the redder when brighter (RWB) trend at the low flux state, but turns to the bluer when brighter (BWB) trend at the high flux state. While, the $\gamma$-ray spectral index always shows the softer when brighter (SWB) trend. We propose that such complex variation trends can be explained by the increasing jet component with two constant components. The  {optical} PD flares and  {optical} flux flares are not synchronous. It seems that one flux peak are sandwiched by two PD peaks, which have inverse rotation trajectories in the $qu$ plane. The helical jet model can schematically show these characteristics of polarization with fine tuned parameters. The change of viewing angle is suggested to be the primary variable which lead all these variations, although other possibilities like the shock in jet model or the hadronic model are not excluded completely.

\end{abstract}




\section{Introduction} \label{sec:intro}

Blazars are of one subclass of active galactic nuclei (AGNs), which have relativistic jets directed  towards the earth \citep{Urry:1995}. The most remarkable feature of blazars is their violent variability in the broad-band wavelengths. The spectral energy distribution (SED) of blazars exhibits two bumps, which are interpreted as the synchrotron and inverse Compton scattering processes in the leptonic model \citep{Konigl:1981,Sikora:1994,Celotti:2008,Sikora:2009}, or the synchrotron radiation of electrons and protons in the hadronic model \citep{Mucker:2003,Bottcher:2013}. Except several nearby sources, blazars could only be identified as point-like objects in the  optical and $\gamma$-ray bands. The location of the optical and $\gamma$-ray emitting regions could not be resolved directly. The time lag analysis between time series of multi frequencies can help to answer this question \citep{Kudryavtseva:2011,Max:2014a}. The color behavior in variation is one popular method to investigate the variation mechanism. Recently, it was suggested that the correlation between the polarization degree (PD) and optical fluxes can also help to figure out the primary variation mechanism \citep{Shao:2019}. Besides, the Stokes parameter can provide us with double information than the flux in principle, which can provide further evidences to diagnose the variation mechanism. Thus, a comprehensive study of various aspects of variation phenomena can promote us to understand the nature of blazars essentially.

Blazar AO $0235+164$ (IAU name J$0238+1636$) was identified  as a BL Lac object \citep{Spinrad:1975}. Its J$2000$ coordinate is R.A.=02:38:38.9, decl.=+16:36:59 \citep{Johnston:1995}, and the redshift is $z=0.94$ \citep{Cohen:1987}.
Historically, this target has been investigated from various aspects. This target was described as a compact and violently variable source \citep{Davis:1982}. In 1975, its optical brightness increased a hundred-fold in a few weeks \citep{Rieke:1976,Ledden:1976}.  The target showed strong $\gamma$-ray flares in September and October 2008 \citep{Corbel:2008,Foschini:2008}, as well as in August 2015 \citep{Ciprini:2015}.  A good review of the monitoring for this target was given by \citet{Ackermann:2012}.
Its long term light curves of the optical and radio have been monitored, and the periodicity has been discussed \citep{Jenkins:1996,Villata:1999,Raiteri:2001,Ostorero:2004,Liu:2006,Wang:2014}.
\citet{Raiteri:2001} revealed that the outbursts of this source indicate a period of $\sim 5.7$ years.
\citet{Ostorero:2004} found that the periodicity of the main optical and radio outbursts and the SED can be well explained by the helical jet model \citep{Villata:1999}. It was also proposed that the periodicity may be caused by the oscillatory accretion of the super massive black hole binary (SMBHB) \citep{Liu:2006}.
The correlations between different wavelengths were widely discussed \citep{Ledden:1976,Macleod:1976,Kraus:1999,Raiteri:2001,Chen:2002,Agudo:2013,Volvach:2015,Fan:2017,Hagen:2018,Kutkin:2018}.
Using the discrete correlation function (DCF) analysis,  \citet{Chen:2002} found that the variation of the short wavelength comes earlier  than that of the long one.   \citet{Agudo:2013} proposed that the emitting region of high energy photons can be inferred from the relative position of the features in the Very Long Baseline Interferometry (VLBI) images. \citet{Fan:2017} indicated that the optical variation leads the radio  for $23\pm 13$ days. \citet{Hagen:2018} presented the multi-wavelength monitoring results from 2007 to 2015, and found that there was no time lag between the optical and $\gamma$-ray, indicating that the optical and $\gamma$-ray were radiated from the same region.  Using both the time lags and radio core size measurements, \citet{Kutkin:2018} derived the distance between radio core and the jet apex, estimated the jet parameters, and indicated that the outflow is bend.

The variation of polarization and its correlation with optical flux can constrain the variation mechanism.
\citet{Cellone:2007} found that total flux of the target is correlated neither with the color nor with the  {PD}.  A model of transverse shock propagating along the jet was proposed to explain the correlation between the flux and PD in the outburst during 2006 December \citep{Hagen:2008}. Due to the correspondence between a series of optical PD flares and  a pronounced maximum polarization of a  superluminal feature at 7 mm, \citet{Agudo:2011} suggested that the ordering degree of the magnetic field fluctuates rapidly. \citet{Cellone:2007} found a trend that the color turns redder when the PD is higher. \citet{Sasada:2011} reported that the short flares in 2008 and 2009 show a bluer when brighter (BWB) trend. They also found a significant positive correlation between the amplitudes of the flux density and the PD in short flares. \citet{Ackermann:2012} found that strong $\gamma$-ray flares are accompanied by the increase in optical PD. Recently, \citet{Shao:2019} found that there is a significant correlation between PD and optical flux density for PKS 1502+106, which can be explained if the variation is due to the change of the viewing angle. This provides us with more clues on the variation mechanism for AO 0235+164.

The difficulty to understand the variation of blazars is due to that there are many emission components and variable factors in AGNs.  A comprehensive analysis of various  phenomena of variation is necessary to pin down the variables. In this paper, we collect the long term data of AO 0235+164, including the $\gamma$-ray, optical and radio, as well as the spectra and polarization. We analyze the correlation between different bands to locate the emitting regions. We study the variable behaviors of color and spectral index of $\gamma$-ray to  figure out the primary variation mechanism. The optical polarization data, including the time series of PD  and trajectories of $(q,u)$, constrain the
variation mechanism in a further step. Our main conclusion is that the helical jet model can roughly explain all these variation phenomena of AO 0235+164, and the viewing angle is the primary variable. The shock in jet model is not excluded completely.
This paper is organized as follows. In Section \ref{Sec:data}, nearly nine years archival data are collected and reduced. In Section \ref{Sec:lag}, the LCCF and time lag analysis is presented. Then, the location of the $\gamma$-ray and optical emitting regions is derived. In Section \ref{Sec:variable}, the spectral index behaviors and variation of polarizations are manifested. The helical jet model is simulated to explain the observational phenomena. At last, our conclusion is given in Section \ref{Sec:conclusion}.

\section{Data collection and reduction} \label{Sec:data}
In this work, we collected the  multi-wavelength data of AO 0235+164 from the blazar monitoring programs, which are retrieved from the {\it Fermi Science Support Center} (FSSC)\footnote{\url{https://fermi.gsfc.nasa.gov/ssc}}, the Steward Observatory\footnote{\url{http://james.as.arizona.edu/~psmith/Fermi}.}, the Small and Moderate Aperture Research Telescope System (SMARTS)\footnote{\url{http://www.astro.yale.edu/smarts/glast/}.}, as well as the Owen Valley Radio Observatory (OVRO)\footnote{\url{http://www.astro.caltech.edu/ovroblazars/}.} \citep{Atwood:2009,Smith:2009,Bonning:2012,Richard:2011}.

{\bf Fermi LAT data}
The {\it Fermi} large area telescope (LAT) data of AO 0235+164 which spans nearly nine years from 2008 August 4 to 2017 June 30 are downloaded. The energy of $\gamma$-ray photons is selected to be in the range of  {$0.1-300$ GeV}.  The region of interest (ROI) centered on the target has a radius of $15^{\circ}$. The {\it Fermitools} (version 1.0.1) was installed within the {Conda package manager}. We use the {\it unbinned likelihood} method to reduce the $\gamma$-ray data \citep{Abdo:2009}. In the standard pipeline, the XML model files are generated by using {\bf make3FGLxml.py}. The instrument response function is {\bf P8R2\_SOURCE\_V6}. The Galactic diffusion background template model {\bf GLL\_IEM\_V06.fit} and the extragalactic isotropic dispersion background {\bf iso\_P8R2\_SOURCE\_V6\_v06.txt} are accounted.
The target AO 0235 is named as `3FGL J0238.6+1636' in the 3FGL catalog, which has a log-parabola spectrum $dN/dE \propto (E/E_b)^{-(\alpha+\beta log(E/E_b))}$ \citep{Acero:2015}.
Parameters of point sources within the radius of $5^{\circ}$ from the center of ROI were left free, while the others are fixed according to the values given in the 3FGL catalog.
To obtain the spectral indices of $\gamma$-ray, we divide the energy range of $0.1-219$ GeV  into seven logarithmically equal bins,  and set the time bin as 4 days.
%This yields the $\gamma$-ray light curves at different energy intervals.
We perform the SED fitting by using the $\gamma$-ray fluxes of different energy bins in one time bin.  {To obtain the high quality spectral index data, we
set the threshold of test statistics (TS) value to be 10. Fluxes with TS less than 10 will be ignored in the SED fitting.} The fitting function $\log F_{\gamma}(E) = \alpha \log E + A_0$ is adopted. Thus, one obtains the $\gamma$-ray spectral indices of power law spectra. The variation behavior of spectra  will be studied by pairing the simultaneous flux and spectral index.

{\bf Photometry and polarization data}
The optical photometry and polarization data are taken from the Steward Observatory (SO), which runs the ground-based  monitoring program to support the {\it Fermi} LAT program \citep{Smith:2009}. We collect the {\it V}-band  and {\it R}-band photometry data from 2008 October 4 to 2018 February 12. Considering the moon phases and weather conditions, as well as other competitive times, the observing cadence of the data is uneven. The PD and the Stokes parameters ($q = Q/I$ and $u =U /I$) observed by the SPOL polarimeter are also sorted out, which provide more constraints for the variation models. The interstellar polarization is not considered in the calibration of $q$ and $u$.
We also collect the optical and near-infrared (NIR) data from SMARTS \citep{Bonning:2012}. This monitoring project uses the 1.3-m telescope, which was mounted with the {\it BVRI} and {\it JHK} filters. To study the color index behavior, we consider only $B$, $V$, and $J$ bands data in this work. The time interval of the SMARTS data is from 2008 February 5 to 2015 August 31. The intercalibration of {\it V}-band data from both SMARTS and Steward is performed by the same comparing star in the finding charts. We combine them to produce the {\it V}-band light curve.

\begin{figure*}[htbp]
 \centering
 \includegraphics[scale=0.55]{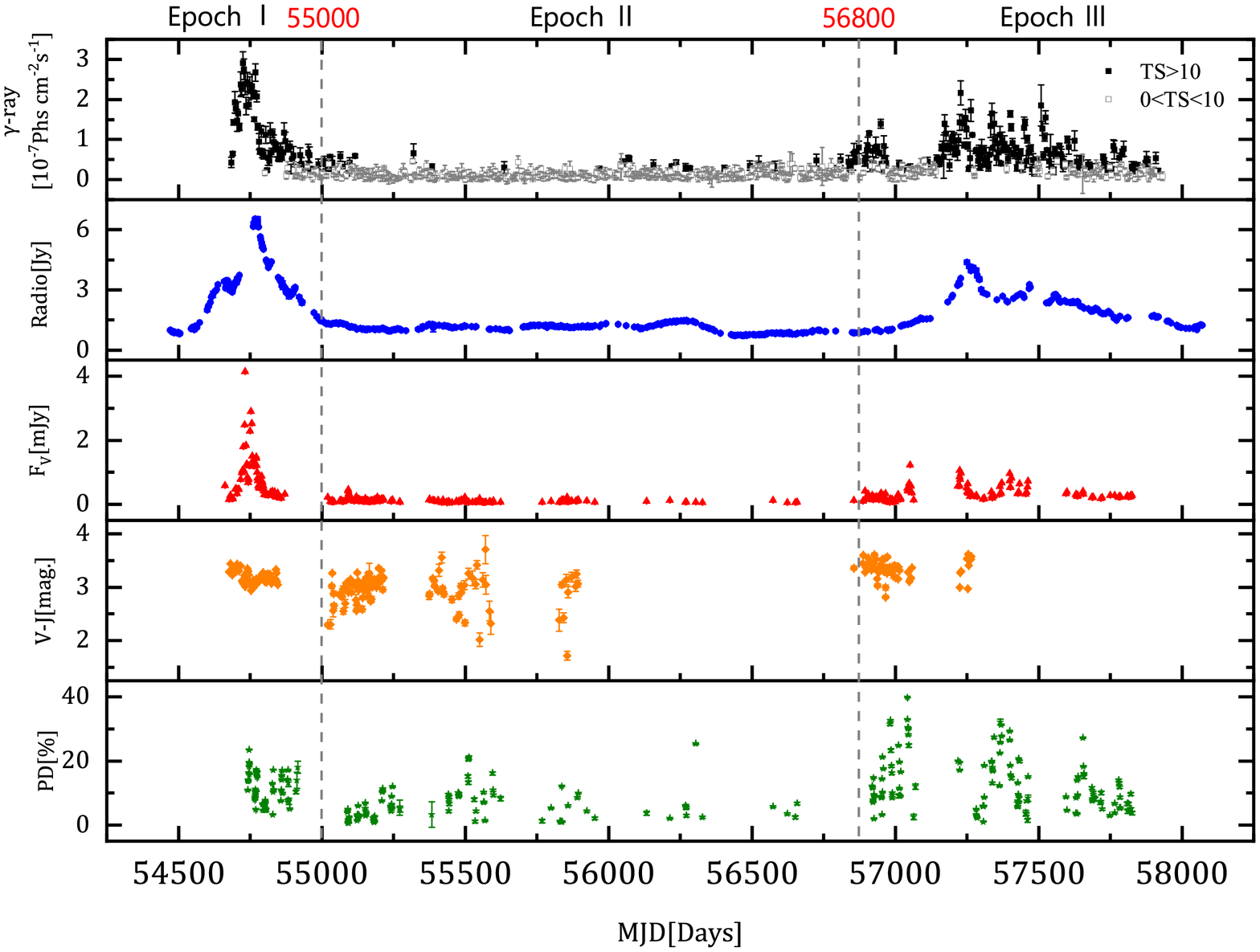}
 \caption{From top to bottom panels, the light curves of $\gamma$-ray in the range of  {$0.3-0.9$ GeV}, radio 15GHz, optical {\it V}-band, {\it V}-{\it J} color index and PD are plotted, respectively. Three evident epochs, namely the flare, the quiescent, and the active states, are separated by grey vertical dash lines at MJD 55000 and 56800, respectively.  {In the top panel, the black squares denote the data of TS$>10$, and the grey empty squares denote the data of $0<$TS$<10$.}}
 \label{LC}
\end{figure*}

{\bf Radio $15$ GHz data}
The 15 GHz radio flux data was obtained from Owen Valley Radio Observatory (OVRO) 40-m monitoring program \citep{Richard:2011}. In this work, the calibrated data span from 2008 June 6 to 2017 November 14. In this time duration, 569 data points are present.
In Figure \ref{LC}, the $\gamma$-ray  in the energy range of  {$0.3 - 0.9$ GeV},  optical {\it V}-band, radio 15 GHz, $V-J$ color index, and PD light curves are presented.

\section{Locations of emitting regions}  \label{Sec:lag}

\subsection{Time lags}

The correlations of multi-wavelength light curves are important to reveal the emission mechanism of blazars. Two methods are often used to calculate the correlation between two time series with uneven samplings, i.e., the discrete correlation function (DCF) \citep{Edelson:1988} and the local cross correlation function (LCCF) \citep{Welsh:1999}.
By comparing the performance of two methods, \citet{Max:2014b} proposed that the LCCF method can manifest physical signals more effectively than the DCF method.
In addition, the coefficients of DCF can exceed the range of $[-1, 1]$, make it unapplicable for the standard statistical test \citep{White:1994}. The coefficients of LCCF are in the range of $[-1, 1]$, which is a good property for the significance level estimation. Thus, we use the LCCF to calculate the correlations in this work.
In order to estimate the significance levels of signals, the Monte Carlo (MC) simulation is used \citep{Shao:2019}. First, $10^4$ artificial radio light curves are simulated by using the Timmer-K\"{o}nig algorithm  (TK95) \citep{Timmer:1995}. For AO 0235+164, the slope of the power spectral density (PSD)  {for the simulated radio light curve} is set to be $\beta$ = 2.3 referring to \citet{Max:2014a}.  Each simulated light curve contains  {10000 data points and the time bin is 1 day}. Since the observed radio light curve is unevenly sampled (US), we extract a  {subset time series which has the exact same samplings of the observed one.}
Secondly, LCCFs between the simulated US time series and the observed $\gamma$-ray (or optical {\it V}-band) light curve are calculated at each lag bin. Then, the 1$\sigma$, 2$\sigma$, and 3$\sigma$ significance levels corresponding to the chance probability of $68.26\%$, $95.45\%$ and $99.73\%$ can be obtained.
Having the significance levels, the signals of lag and its 1$\sigma$ standard deviation estimation are based on the model independent Monte Carlo method proposed by \citet{Peterson:1998}.
This method considers both the flux randomization (FR) and the random subset selection (RSS) processes \citep{Peterson:1998,Larsson:2012}.
Two kinds of time lags can be calculated, $\tau_p$ and $\tau_c$. $\tau_p$ is the lag for the highest peak of LCCF, while $\tau_c$ is the centroid lag defined as $\tau_{c}\equiv \sum_{i} \tau_{i} C_{i} / \sum_{i} C_{i}$  , where $C_i$ is the correlation coefficient satisfying $C_i > 0.8 {\rm LCCF}$($\tau_p$) \citep{Shao:2019}. We repeated $10^4$ times to obtain the distribution of $\tau_p$ and $\tau_c$.  The error of time lag is taken as 1$\sigma$ standard deviation.

\begin{figure*}[ht]
%   \begin{tabular}
    \begin{minipage}[t]{0.48\linewidth}
        \centerline{\includegraphics[scale=0.35]{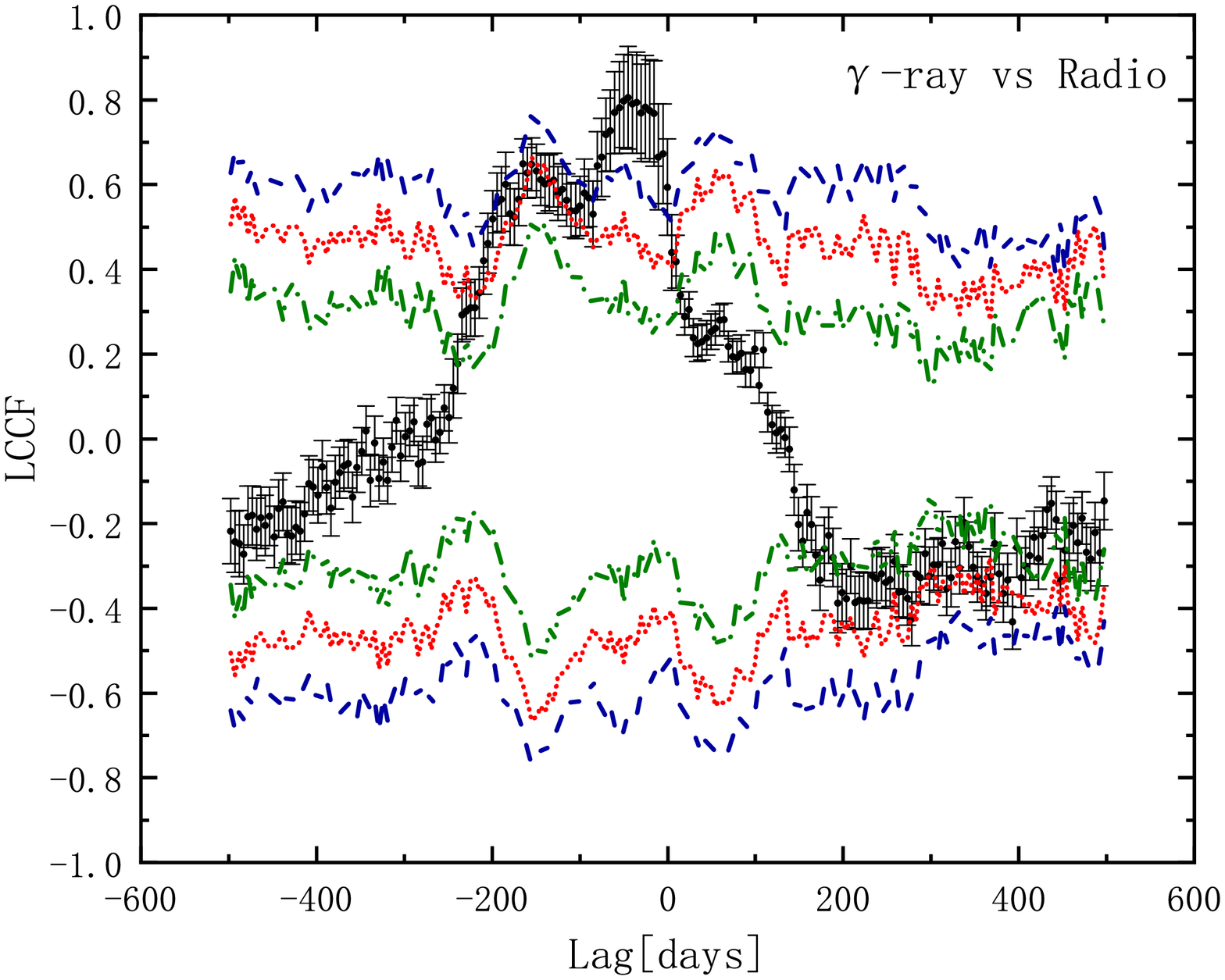}}
           \end{minipage}%
    \begin{minipage}[t]{0.48\linewidth}
        \centerline{\includegraphics[scale=0.35]{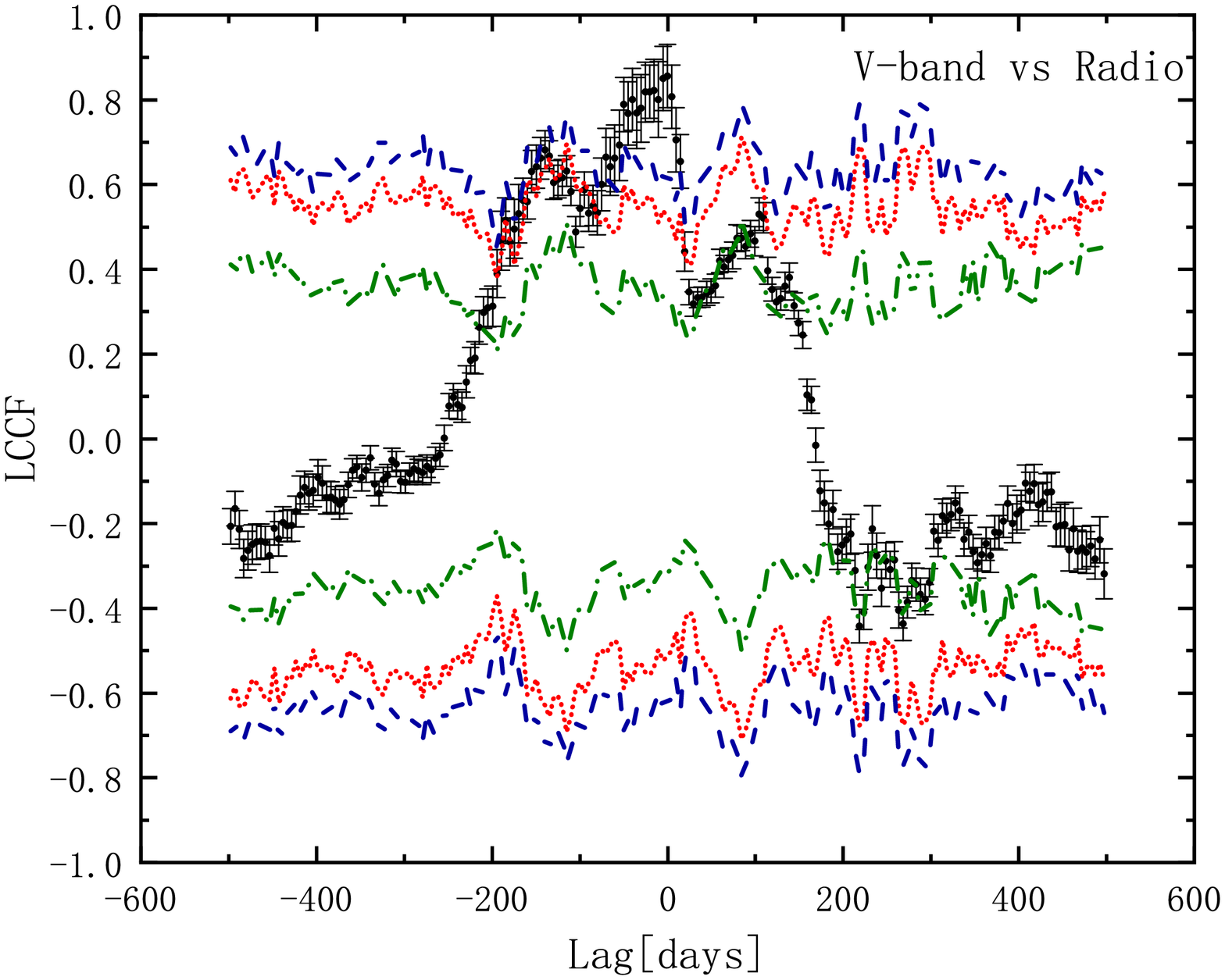}}
    \end{minipage}
    \caption{Left and right panels present the plots of LCCF results of $\gamma$-ray  {(TS$>$10)} versus radio (15 GHz) and optical {\it V}-band flux versus radio, respectively. Black dots with error bars denote correlation values. The olive dashed dotted, red dotted and royal dashed lines denote the $1\sigma$, $2\sigma$ and $3\sigma$ significance levels, respectively. The negative lag indicates that the former leads to the latter. }
    \label{Fig:LCCF}
\end{figure*}

In Figure \ref{Fig:LCCF}, LCCFs of the $\gamma$-ray and optical {\it V}-band versus radio 15 GHz are plotted in the left and right panel, respectively.  We use the $\gamma$-ray light curve in the energy interval of  {$0.3-0.9$ GeV to do the LCCF calculation. This light curve has the TS threshold of $10$, and $68\%$ of data points are ignored. Large gaps appear in the reduced $\gamma$-ray light curve. If the TS threshold is taken to be $0$, only 14$\%$ of data are ignored. The lag result is almost invariant under different TS thresholds, but the fluctuation of significance estimation is reduced with the smaller TS threshold.} The lag range is [-500, 500], and the lag bin is 5 days.
In the left panel of Figure \ref{Fig:LCCF}, the peak of LCCF between the $\gamma$-ray and radio light curves is beyond 3$\sigma$ significance level.  The $\gamma$-ray leads radio with   {$\tau_p=-45.3^{+14.9}_{-5.0}$ and $\tau_c=-45.4^{+7.3}_{-14.4}$} days. \citet{Max:2014a} presented that the $\gamma$-ray leads radio with $150\pm 8$ days with $99.99\%$ significance. However, there are three peaks in the plot of LCCF (see the top right panel in Figure 1 in the reference). Among them, the most significant one locates at $-30$ days, which roughly agrees with our result. The flares in the light curve can significantly affect the correlation results. In their work, only four years $\gamma$-ray data has been used, which contains only the flare state and quiescent state. Our $\gamma$-ray data covers more than eight years, and the active state (MJD 56800 to MJD 58000) has been recorded. The long term duration and complex states will enhance the significance of our result. In the right panel of Figure \ref{Fig:LCCF}, the LCCF between the {\it V}-band and radio light curves also shows one peak beyond the 3$\sigma$ significance level. The optical {\it V}-band leads the radio with $\tau_p=0.5^{+5.0}_{-0.0}$ and $\tau_c=30.1 ^{+18.1}_{-9.8}$ days.  \citet{Raiteri:2001} indicated that the time lags between optical and radio vary for different outbursts. \citet{Fan:2017} used DCF to show that the optical lead the radio with $23\pm13$ days. These results did not conflict with our result.
In Table \ref{Table:Lag}, one also notes that the $\tau_c$ results are consistent for the three relative time lags, while $\tau_p$ results are inconsistent. The difference between $\tau_p$ and $\tau_c$ may be due to that the observation gap in the optical and radio monitoring programs can lead to the deformation of LCCF to some extent \citep{Shao:2019}. It is also possible that the propagation of disturbance from optical to radio varies each time, which leads the uncertainty of the lag.

\begin{figure}
\center
\includegraphics[scale=0.5]{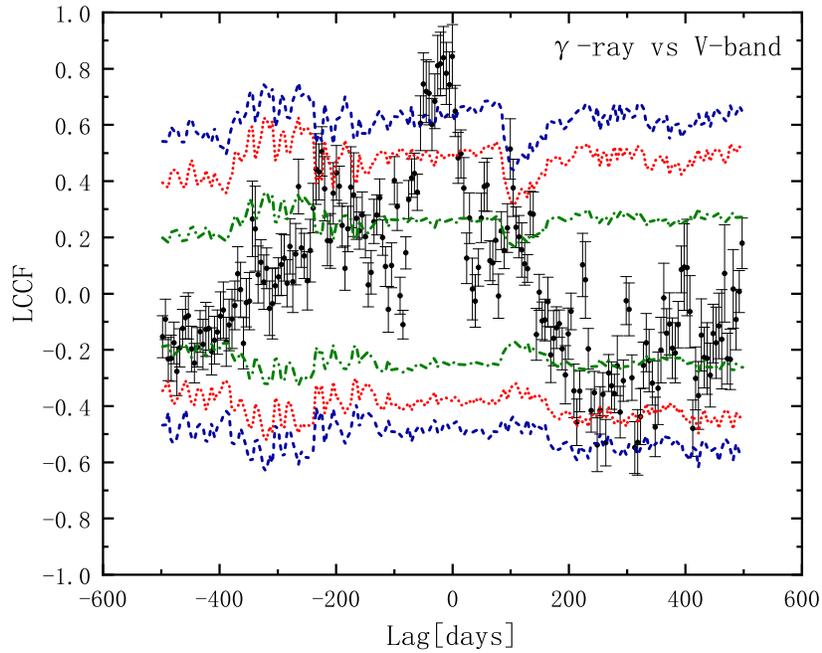}
\caption{The LCCF between $\gamma$-ray  {(TS$>$10)} and {\it V}-band light curve is plotted. The markers for the significance levels are the same as Figure \ref{Fig:LCCF}.}
\label{Fig:LCCFGO}
\end{figure}

In Figure \ref{Fig:LCCFGO}, the LCCF between $\gamma$-ray and {\it V}-band is plotted. In the MC simulation of significance levels,  we use TK95 algorithm and set $\beta=1.0$ (referred to \citealt{Max:2014a}) to produce $10^4$ artificial $\gamma$-ray light curves. Each curve contains $10000$ days with 1 day time bin. To mimic the reduced $\gamma$-ray light curve, we rebin the artificial curves with $4$ days time interval, and take exactly the same samplings with the {$\gamma$-ray light curve with TS$>10$ .}  The FR/RSS procedures present  { $\tau_p=-5.5^{+5.0}_{-14.9}$ and $\tau_c=-15.1^{+4.7}_{-5.8}$} days, respectively. From the result of $\tau_c$, it seems that the $\gamma$-ray leads the {\it V}-band. However, the $\gamma$-ray and optical emissions are possibly simultaneous within the uncertainty of $\tau_p$. \citet{Hagen:2018} claimed that there are no time delay between the $\gamma$-ray and optical light curves, but no correlation analysis was performed to exhibit that. With the current precision of samplings, the optical emitting region probably is the same as $\gamma$-ray emitting region. More intensively samplings help to improve the precision of LCCF results.

%The most significant peak indicates that there is most probably no time delay between the $\gamma$-ray and optical {\it V}-band.

\begin{deluxetable}{cccccccc}
\tabletypesize{\scriptsize}
%\rotate
\tablecaption{Time lags \label{Table:Lag}}
\tablewidth{0pt}
\tablehead{
\colhead{Time lags} & \colhead{$\gamma$-ray vs radio} & \colhead{{\it V}-band vs radio} & \colhead{$\gamma$-ray vs {\it V}-band} }
\startdata
$\tau_p$ (days)&   {$-45.3^{+14.9}_{-5.0}$} & $-0.5^{+0}_{-5.0}$ &    {$-5.5^{+5.0}_{-14.9}$} \\
$\tau_c$ (days)&   {$-45.4^{+7.3}_{-14.4}$} & $-30.1^{+9.8}_{-18.1}$ &    {$-15.1^{+4.7}_{-5.8}$}    \\
\enddata
\tablecomments{Here $\tau_p$ and $\tau_c$ denote the centroid and peak time lags (in unit of days), respectively. The negative lags indicate that the former lead the latter.}
\end{deluxetable}

\subsection{Location of emitting regions}

The time lag between $\gamma$-ray and radio can be well explained by the jet model, i.e., the disturbance propagates along the jet, and $\gamma$-rays and radio are radiated in the upstream and downstream, respectively.
The distance between emitting regions of different bands is given by  \citep{Kudryavtseva:2011,Max:2014a}
\begin{equation} \label{Eq:d}
\Delta \mathrm{D}=\frac{\beta_{\rm app} c \Delta T}{(1+z) \sin \theta},
\end{equation}
where $\beta_{\rm app}$ is the apparent velocity in the observer frame, $c$ is the speed of light, $\Delta T$ is the time lag ($\tau_p$ or $\tau_c$) between different bands. For this target, the redshift is  $z=0.94$ \citep{Cohen:1987}. \citet{Lister:2009} stated that $\beta_{\rm app}$ was not available due to the poor resolution of VLBI images at 15 GHz. One can take advantage of the VLBI 43 GHz data to estimate $\beta_{\rm app}$.  \citet{Jorstad:2017} presented the apparent velocity $\beta_{\rm app}$ of three jet components of AO 0235+164, which are $26.27\pm1.67$, $13.39\pm1.47$, and $3.05\pm0.31$, respectively. \citet{Kutkin:2018}  {incorporated the birth date, time-resolved feature size, and distance from the core to conclude that the true physical component indicates an apparent velocity $\beta_{\rm app}=10$}.
For the viewing angle $\theta$, \citet{Hovatta:2009} derived that $\theta=0.4^{\circ}$ based on $\beta_{\rm app}=2$.  {We take $\beta_{\rm app}=10$ to calculate the distance, and
derive the viewing angle to be $\theta=1.7^{\circ}$. }

The relative distances between emitting regions of different bands, i.e., $D_p$ and $D_c$, are calculated according to Equation (\ref{Eq:d}) by inserting $\tau_p$ and $\tau_c$, respectively. The results are summarized in Table \ref{tab:D}. The $\gamma$-ray emitting region locates at a distance about   {$7$} parsecs (pc) in the upstream of the core region of 15 GHz. There are large uncertainties of the time lag between {\it V}-band and radio due to unevenly sampled data. We tend to adopt the relative significant result between $\gamma$-ray and {\it V}-band to determine the optical emitting region, i.e., the $\gamma$-ray and the optical emitting regions coincide within error.

To obtain the locations of emitting regions, we need to determine the distance between the core region of 15 GHz and the jet base, which is denoted as $r_{\rm core}$. Two methods can be applied to derive $r_{\rm core}$. The first method considers the geometrical relation for a conical jet, which is given as \citep{Hirotani:2005}
\begin{equation} \label{Eq:r_c1}
r_{\text {core }}=\frac{r_{\perp}}{\varphi}=\frac{0.5 \theta_{\mathrm{d}} d_{L}}{(1+z)^{2} \varphi},
\end{equation}
where $r_{\perp}$ is the jet transverse size at the core position, $\theta_d$ is the angular diameter at a certain frequency, $\varphi$ is the half-opening angle of the jet.
Following \citet{Kutkin:2018}, the half-opening angle is $\varphi \approx 1^{\circ}$, and the core size of 15 GHz  is $\theta_d \approx 0.125$ mas. The redshift $z=0.94 $ indicates that the luminosity distance of the target is $d_L =6142$ Mpc \citep{Cohen:1987}. With these parameters, $r_{\rm core, 15 GHz}$ is derived to be $29$ {\rm pc}.
 Core-shift measurement is the second commonly used method to derive $r_{\rm core}$ \citep{Konigl:1981,Lobanov:1998,Hirotani:2005}. Considering the opacity, the positions of photospheres will shift systematically toward the downstream of the jet as the frequency decreases, which can be observed by VLBI images at millisecond resolution. $r_{\rm core}$ at the frequency $\nu$ can be derived as \citep{Lobanov:1998,Hirotani:2005}
\begin{equation}
r_{\text {core }}(\nu)=\frac{\Omega_{r \nu}}{\sin \theta} \nu^{-1/k_r}, \qquad \Omega_{r \nu} \equiv 4.85 \times 10^{-9} \frac{\Delta r_{\nu_{1} \nu_{2}} d_{L}}{(1+z)^{2}} \frac{\nu_{1}^{1 / k_{r}} \nu_{2}^{1 / k_{r}}}{\nu_{2}^{1 / k_{r}}-\nu_{1}^{1 / k_{r}}} \mathrm{pc} \cdot \mathrm{GHz}.
\end{equation}
where $\Omega_{r\nu}$ is the core position offset.
\citet{Kutkin:2018} presented  $\Delta r_{15 \rm GHz, 5 GHz}=0.1{\rm mas}$ and $k_r=1.25$. With these parameters, one obtains  $\Omega_{r\nu}=4.9\ {\rm pc}$ GHz and $r_{\rm core, 15 GHz}\approx 18.9$ pc.
Based on the geometrical method, \citet{Max:2014a} have estimated $r_{\rm core, 15 GHz} \gtrsim 23 \pm 6$ pc by using averaged core angular size of multiple epochs. Our results of $r_{\rm core, 15Ghz}$ are consistent with
the constraint given by \citet{Max:2014a}. Taking $r_{\rm core, 15 GHz}\sim 18.9$ pc,  the location of  the $\gamma$-ray and optical emitting region is about  {$12$} pc away from the jet base, which is far beyond the BLR region.

\begin{deluxetable}{cccc}
%\tabletypesize{\scriptsize}
\tablecaption{Relative distances \label{tab:D}}
%\tablenum{2}
%\tablewidth{0pt}
\tablehead{
\colhead{Distance} & \colhead{$\gamma$-ray vs Radio} & \colhead{{\it V}-band vs Radio} & \colhead{$\gamma$-ray vs {\it V}-band}}
\startdata
  $D_p$ (pc)&    {$6.6_{-1.7}^{+0.6}$} & $0.1_{-0.0}^{+0.7}$ &    {$0.8_{-0.7}^{+2.1}$}\\
  $D_c$ (pc)&    {$6.6_{-0.1}^{+0.3}$} & $4.4_{-1.4}^{+2.6}$ &    {$2.2_{-0.7}^{+0.9}$}\\
\enddata
\tablecomments{ $D_p$ and $D_c$ (in units of pc) are distances derived according to $\tau_p$ and $\tau_c$ in Table \ref{Table:Lag}, respectively. }
\end{deluxetable}

With the known $\Omega_{r\nu}$, the magnetic field strength and electron number density at 1 pc can be derived as \citep{Hirotani:2005,OSullivan:2009,Kutkin:2018,Jiang:2020}
\begin{align}
B_{1} & \simeq 0.025\left[\frac{\sigma_{\mathrm{rel}} \Omega_{r \nu}^{3k_r}(1+z)^{2}}{\varphi \sin ^{3k_r-1} \theta \delta^{2}}\right]^{1 / 4}, \\
N_{1} & \simeq 3.3\left[\frac{\sigma_{\mathrm{rel}} \Omega_{r \nu}^{3k_r}(1+z)^{2}}{\gamma_{\mathrm{min}}^{2} \varphi \sin ^{3k_r-1} \theta \delta^{2}}\right]^{1/2},
\end{align}
where $\gamma_{\rm min}$ is the minimum Lorentz factor of radiative electrons,
and $\sigma_{\rm rel}$ is the ratio of the magnetic field energy density to the non-thermal particle energy density.  Setting $\sigma_{\rm rel}=1$ and $\Omega_{r \nu}=4.9\ {\rm pc\ GHz}$, one obtains two parameters directly, i.e. $B_1 \approx 0.97$ Gauss and $N_1 \approx 5012 \gamma_{\rm min}^{-1} {\rm cm}^{-3}$.
Considering the matter and fields distribution in jet, magnetic field strength and number density of electrons follows the scaling law $B = B_1 (r/r_1)^{-m}$ and $N=N_1(r/r_1)^{-n}$. Using both the core-shift measure and time lags for radio frequencies, \citet{Kutkin:2018} obtained that the power law indices are $m = 1.2\pm0.1$ and $n = 2.4 \pm 0.2$. Based on these parameters, the magnetic field and electron number density  in the optical emission region are calculated to be $0.05$ Gauss and $12 \gamma_{min}^{-1} {\rm cm}^{-3}$, respectively.

\section{Variation analysis} \label{Sec:variable}

The variation phenomena of AO 0235+164 are rich. Combining with the color index behavior, the variation of polarization can help us to further constraint the emission mechanism.

\subsection{Variation of spectral index}
\begin{figure}[ht]
        \centerline{\includegraphics[scale=0.9]{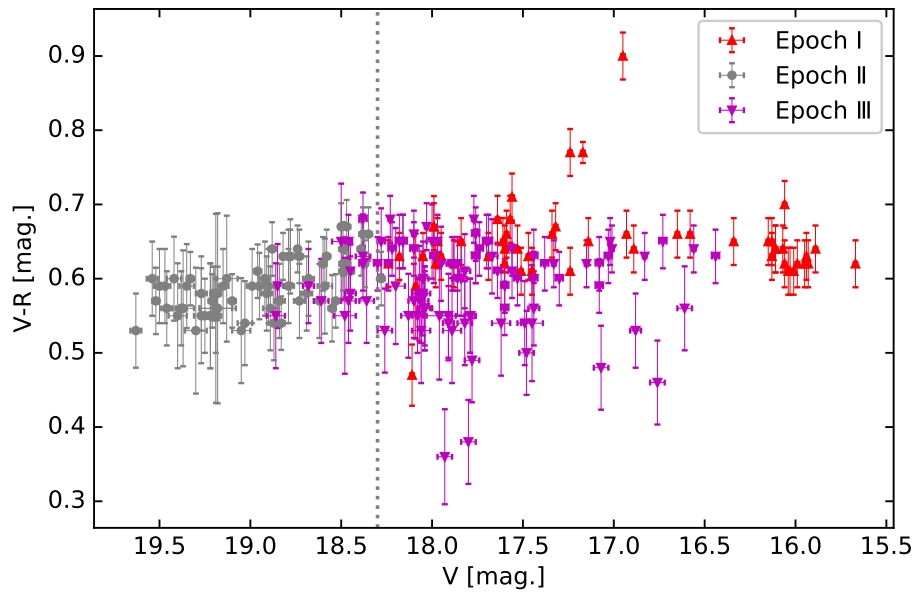}}
%   \end{tabular}
    \caption{$V-R$ color versus {\it V}-band magnitude is plotted. The red triangles, grey circles, and magenta triangles represent data in epoch I, II, and III, respectively. The grey vertical dot line corresponds to the {\it V}=18.3.}\label{V-R}
\end{figure}

\begin{figure}[ht]
%   \begin{tabular}
        \centerline{\includegraphics[scale=0.9]{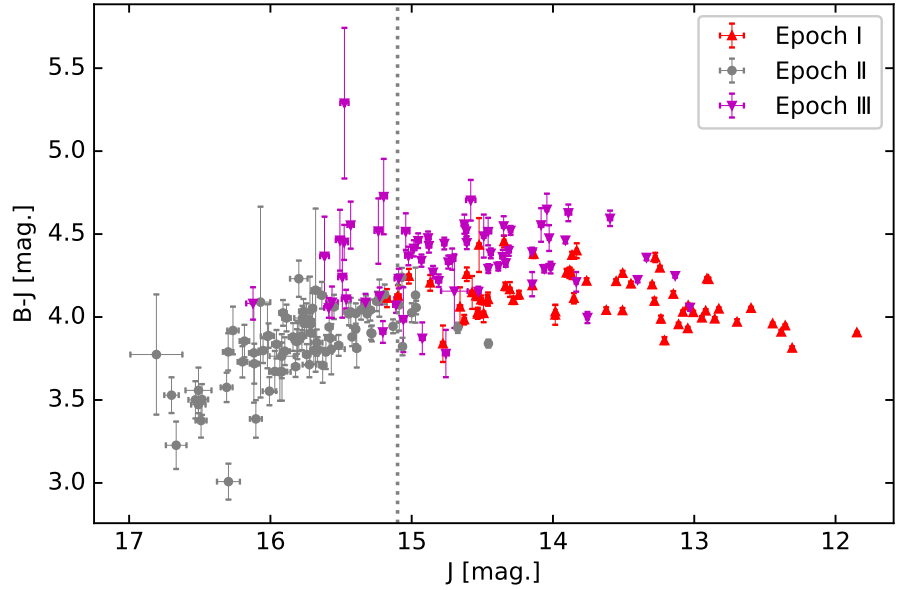}}
         \caption{ $B-J$ color versus $J$-band magnitude is plotted. The marks are the same as in Figure \ref{V-R}. The grey vertical dotted line corresponds to $J$=15.1. The Pearson's $r$  for epoch I, II, and III are 0.37, -0.65, and -0.01, respectively.} \label{V-J}
\end{figure}

\begin{figure}
\begin{center}
  \includegraphics[scale=0.5]{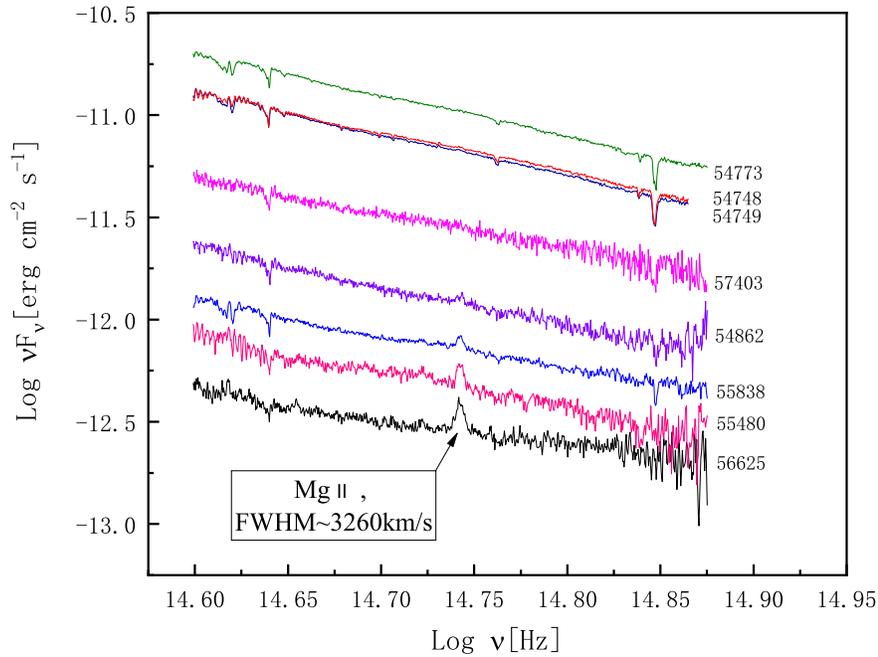}
\caption{The spectra data is taken from Steward observatory. The Galactic interstellar extinction and reddening effects are not corrected. The numbers on the right denote the MJD days.}\label{Fig:Spectrum}
\end{center}
\end{figure}

The color behavior gives us the most direct clues to analyze variation. In Figure \ref{V-R}, the diagram of  $V-R$ color versus {\it V}-band magnitude  is shown.
In epoch II, the variation of color has a weak redder when brighter (RWB) trend (with Pearson's $r_{II} = 0.60$). In epochs I and III, no obvious trends are found ($r_{I}= -0.02, r_{III}= 0.04$).
In Figure \ref{V-J}, the $B-J$ versus $J$-band magnitude is plotted. An obvious RWB trend is evident in the quiescent state. For the flare state (epoch I), the red triangle points indicate a weak bluer-when-brighter (BWB) trend. For the active state, the points are scatter, and the plateau pattern are possibly due to the change of trends. The similar plot was also presented by \citet{Bonning:2012} with 825 days data.  The complex color behavior has been found for CTA 102 (see Figure 3 in \citealt{Raiteri:2017}), which is an FSRQ object. AO 0235+164 was classified as a BL Lac object in \citet{Stein:1976}. However, for this target, \citet{Cohen:1987} and \citet{Raiteri:2007} indicated the broad emission line of Mg II with FWHM 3100 and 3500 km s$^{-1}$, respectively. In Figure \ref{Fig:Spectrum}, we plot the spectra observed by SO. It is evident the broad line of Mg II with FWHM 3260 km s$^{-1}$ appears in the quiescent state, and disappears in the flare and active states. \citet{Chen:2001} investigated the spectral variability of this source during two active phases, and revealed that  AO 0235+164 is much more like an FSRQ object in many aspects. Adopting the quasar composite SED from \citep{Elvis:1994}, the numerical fitting  of the broad band SED indicates that the brightness of the accretion disk component is  100 times smaller than the jet component \citep{Ackermann:2012,Raino:2013}.

The spectra in Figure \ref{Fig:Spectrum} also present us an explanation for the complex variation trend of this target. The observed flux contains both the disc component and the jet component.  In the plot, it is evident that the spectral slopes are negative always. So, the jet component dominates over the disk one even in the quiescent state. The peak frequency of the synchrotron component should be less than $10^{14.6}$ Hz. As the jet component increases,
a RWB trend will be produced.  This mechanism has been suggested to explain the RWB trend for 3C454.3 by \citep{Villata:2006}. When the $J$-band magnitude is less than 15, the $B-J$ color shows an evident BWB trend. In Figure \ref{Fig:Spectrum}, the two lines at MJD 54748 and 54749 show the redder when fainter trend (BWB trend in statistic) in the flux decaying phase.

Two mechanisms can account for the BWB trend, i.e., the geometrical model \citep{Raiteri:2006,Villata:2009,Raiteri:2017} and the shock in jet model \citep{Kirk:1998}.  For this target, many literatures investigated the periodicity \citep{Jenkins:1996,Raiteri:2001,Ostorero:2004,Liu:2006,Wang:2014,Volvach:2015,Fan:2017}.
The helical path in jet model was proposed to explain the periodicity \citep{Ostorero:2004,Raiteri:2006,Villata:2009,Volvach:2015}. The viewing angle between the moving direction and our line of sight varies along the helical trajectory, which results in the variation of Doppler factor. For a decreasing viewing angle, the peak frequency  ($\nu_p \propto \delta$) will shift to higher frequency in the observing frame, and the spectral index of the observing frequency will  increase from negative (when  $\nu_p < \nu_{\rm obs}$) to positive (when  $\nu_p > \nu_{\rm obs}$). The flux also varies as $F_{\nu} \propto \delta^3 F'_{\nu}$. This will produce a BWB trend naturally. For the shock in jet model, \citet{Kirk:1998} presented that the variation shows the BWB trend when the electronic cooling time scale is larger than the accelerating time scale, and shows the RWB trend when the two time scales are approximately the same. The shock in jet model can not be ruled out by the current analysis.

\begin{figure}[ht]
        \centerline{\includegraphics[scale=0.8]{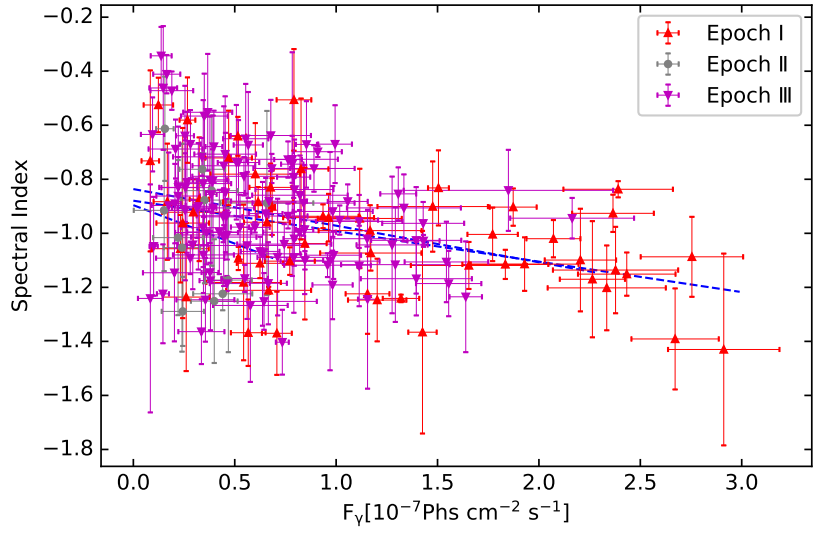}}
         \caption{The $\gamma$-ray spectral index versus $F_{\gamma}$ is plotted. The Pearson's $r$  for epoch I, II, and III are -0.42, -0.20, and -0.27, respectively.} \label{SI}
\end{figure}

\begin{figure}[ht]
%   \begin{tabular}
        \centerline{\includegraphics[scale=0.8]{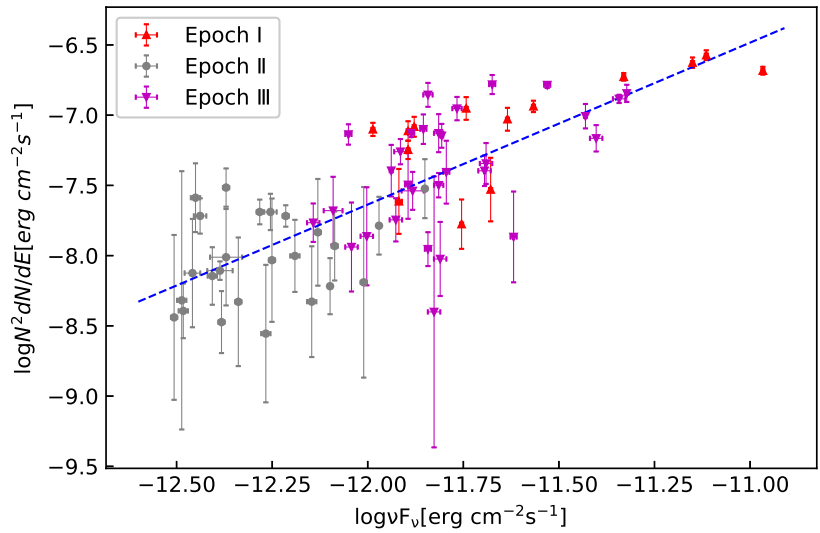}}
         \caption{The log-log diagram of $\gamma$-ray versus {\it V}-band fluxes are plotted. The blue dash line is the linear fitting results with the slope $1.15\pm 0.12$} \label{Fig:OptGamma}
\end{figure}

The spectral index of the $\gamma$-ray versus the $\gamma$-ray flux is plotted in Figure \ref{SI}. In all three epochs, the spectral index decreases as the flux increases. The Pearson's $r$  for epoch I, II, and III are -0.42, -0.20, and -0.27, respectively.  This is the softer when brighter (SWB) trend. \citet{Ackermann:2012} also presented this trend in the fitting of SED.  In Figure \ref{Fig:OptGamma}, we also plot the correlation between $\gamma$-ray and optical {\it V}-band fluxes (log-log diagram). There is a strong linear correlation with slope $1.15\pm0.12$.  This slope can help us further constrain the radiation process of $\gamma$-ray and the variation mechanism according to the theory (see Table 3 in \citealt{Shao:2019}). For the SSC process, the predicted slopes for variables of $N$ (number density), $B$ (magnetic filed) and $\delta$ (Doppler factor) are 2, 1, and $[0.49,1.11]$, respectively. For the EC process, the predicted slopes for variables of $N$, and $\delta$ are 1 and $(0.68,1.57)$, respectively. The case of EC with variable of $B$ can be excluded, since no correlation can be observed for this case. The SSC process is not favored, since output energy of $\gamma$-ray should be smaller than that of optical in SSC. \citet{Ackermann:2012} found that EC process with seed photons from dust torus can fit the SED well, and EC with seed photons from BLR can not reproduce the observed GeV $\gamma$-ray.  We obtained that the emitting region of $\gamma$-ray is far beyond the BLR, which agrees with the EC with seed photons from torus scenario.

For EC, both variables of $N$ and $\delta$ can lead the slope in Figure \ref{Fig:OptGamma}. We need to use the variation trend to further constrain the variation mechanism. The interesting thing is that the optical and $\gamma$-ray fluxes are strongly correlated, but the BWB trend (optical) and the SWB ($\gamma$-ray) trend coexist. In the lepton models, the variation trends should be the same for {\it V}-band and $\gamma$-ray, since the radiative particles are the same population, and both two bands have the negative spectral indices. However, if the spectrum of radiative particles are evolved during the flare,
a broken power law can be formed \citep{Jiang:2016}. The particles radiating $\gamma$-ray photons and the particles radiating {\it V}-band photons distributing at different energy range may have different variation trend. Such process needs detailed numerical simulation and fine tune of parameters to verify, which seems not natural enough to be true. The hadronic models may provide a way out. \citet{Bottcher:2013} presented that the lepton model can not fit the SED of AO 0235+164 well, since the spectrum of $\gamma$-ray for this target is hard, while the IR-optical-UV continuum has a very steep spectrum. Then, it was proposed that the synchrotron of proton can produce the hard spectrum of $\gamma$-ray,  while the synchrotron of electron will correspond to the IR-optical-UV continuum. The electrons and protons must be in the same emitting region, the shock acceleration for them should be similar. So, the hadronic model also needs fine tune to explain the reverse trends. \citet{Ackermann:2012} found that the SED of low and high states can be reproduced  by adjusting the viewing angle and keeping the spectrum of electrons. But, the simulated SED of GeV $\gamma$-ray at low flux state is not good to fit the observed hard spectrum. Similar to the RWB trend for optical continuum, we propose that an invariant component with energy higher than GeV, which is most probably produced by the EC with seed photons from BLR (with energy 10 eV), can lead to the SWB trend. At the low flux state, this constant high energy component will make a hard spectrum of $\gamma$-ray.  As the GeV $\gamma$-ray photons (relatively `soft') of jet component dominates, the spectral indices of $\gamma$-ray will become smaller. This proposal can naturally explain the interesting puzzle for AO 0235+164.

\subsection{Variation of polarization}
\begin{figure}[ht]
%   \begin{tabular}
        \centerline{\includegraphics[scale=0.8]{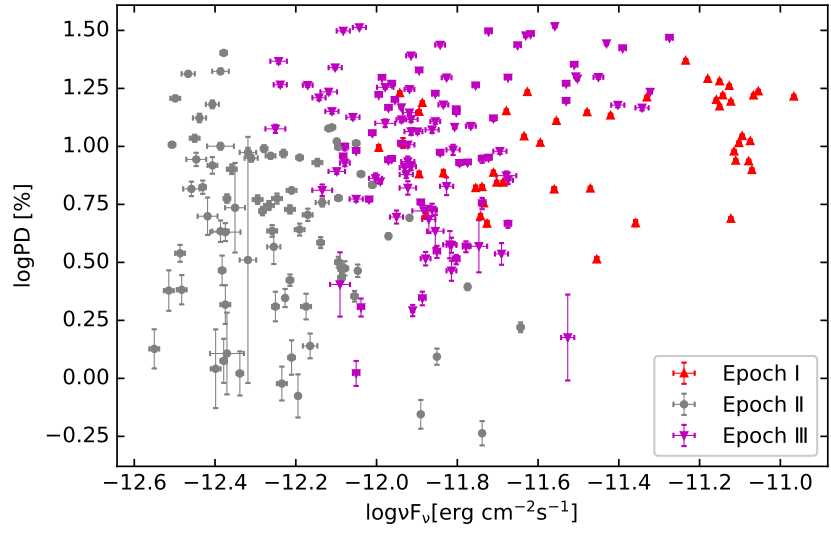}}
         \caption{Distributions of $\log {\rm PD}$  versus $\log \nu F_{\nu}$ of optical {\it V}-band is plotted. } \label{PD}
\end{figure}

\begin{figure}[ht]
%   \begin{tabular}+
        \centerline{\includegraphics[scale=0.8]{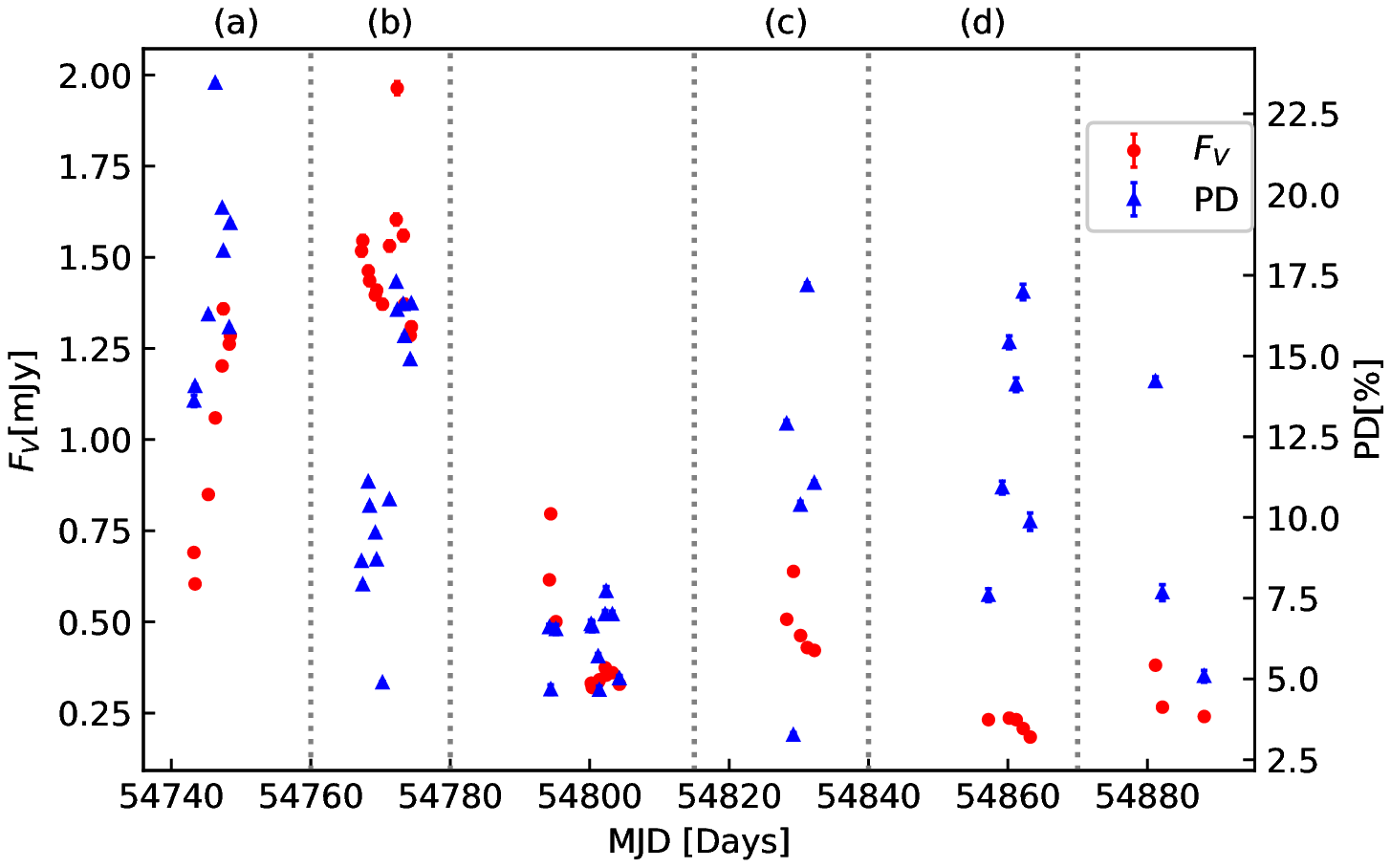}}
        \centerline{\includegraphics[scale=0.8]{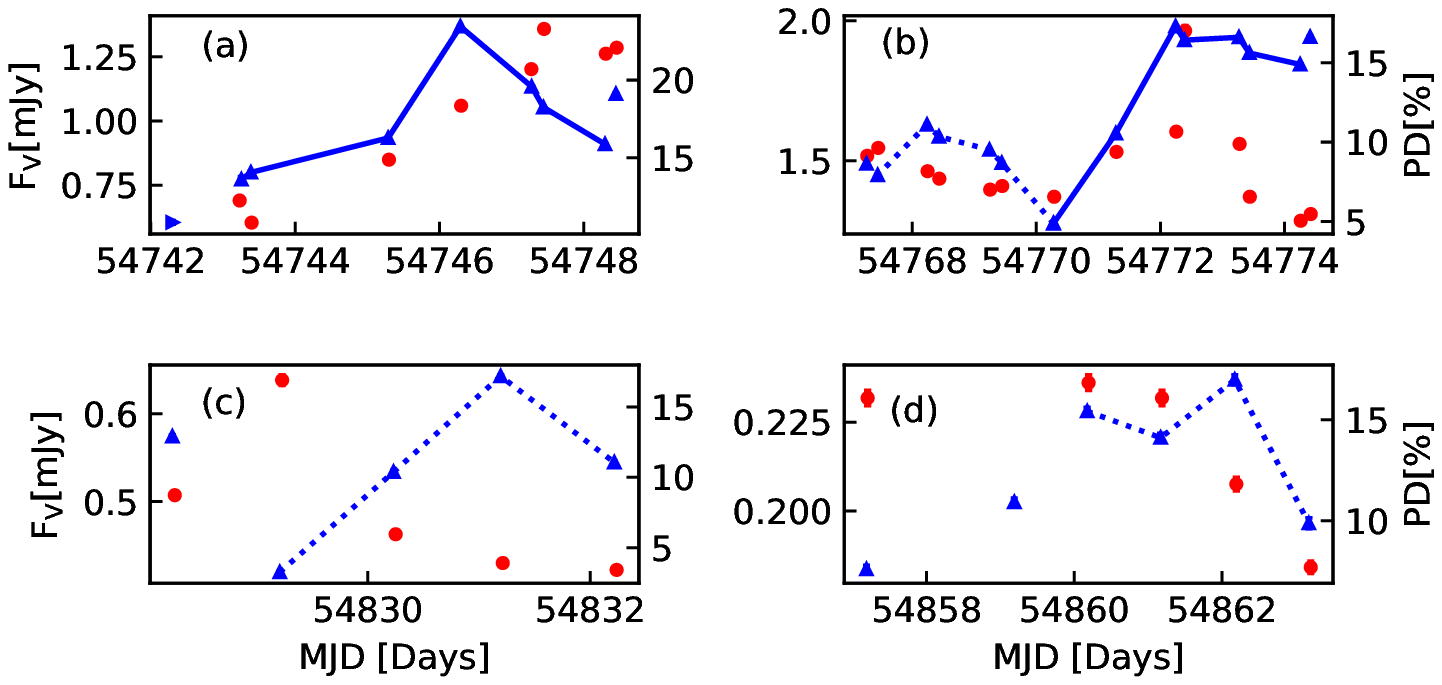}}
\caption{The top panel shows the light curves of the optical {\it V}-band flux (red circle) and PD (blue triangle) in Steward's first cycle. The bottom panels show the data in 4 intensively sampled segments, which are marked with (a), (b), (c), and (d), respectively. The solid line denotes a former PD flare in the increasing phase of one flux flare, while the dotted line denotes a latter PD flare in the decreasing phase of one flux flare.
 }  \label{PDFV}
\end{figure}

\begin{figure}[ht]
%   \begin{tabular}
        \centerline{\includegraphics[scale=0.8]{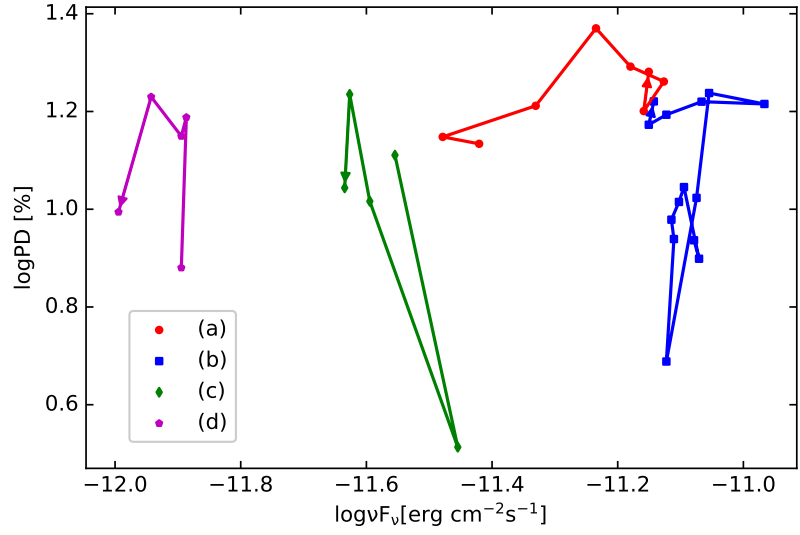}}
         \caption{The plot of log(PD) versus log($\nu F_{\nu}$) for the 4 segments, whose labels are the same as in Figure \ref{PDFV}. The arrows in the end of lines denote the direction of increasing time.} \label{logFPD}
\end{figure}

To further constrain the variation mechanism, we analyze  the polarization data from SO.
In Figure \ref{PD},  we find that $\log {\rm PD}$ has no linear relation with $\log \nu F_{\nu}$ \citep{Ikejiri:2011,Sasada:2011}. \citet{Ikejiri:2011} studied the rotations in the $qu$ plane for a sample of blazars. In which, AO 0235+164 was also studied. It was found that there is a linear correlation between PD and fluxes during the outburst period (Epoch I).  The light curve of the target is similar to PKS 1502+106, which also has three ordered epochs (the flare, the quiescent, and the active states in the Fermi monitored period) \citep{Shao:2019}. However, a significant linear relation between $\log {\rm PD}$ and $\log \nu F_{\nu}$ was found for PKS 1502+106.
In the top panel of Figure \ref{PDFV}, we segment the data of {\it V}-band and polarization observed during Steward's first observing cycle (from 2008 October 3 to 2009 October 26, which is in the flare state, see Figure \ref{LC}) into six periods. Among them, four periods with continuous flare profiles are marked with (a), (b), (c), and (d), respectively. The double logarithmic diagram of  PD versus {\it V}-band flux for the four segments is plotted in Figure \ref{logFPD}. At different flux levels, the polarization varies significantly. It is evident that the superposition of them will produce the uncorrelated relation between PD and the flux. If there is a disk component, the combination of the jet and disk component will diminish the PD in principle. At low flux state, the PD must be small. When the jet component dominates over the disk component, the PD will be high and oscillate. This can naturally explain why the PD is strongly correlated with the flux for PKS 1502+106. However, from the SED of  AO 0235+164 \citep{Ackermann:2012}, one knows that the jet component dominates over the disk component even at low flux levels. This leads to a large variation of PD at the low flux levels for our target. We conclude that the domination of the jet component can explain both the uncorrelation between PD and fluxes and the RWB trend of the color for our target.

The variation of PD at different flares provides us more information to infer the variation mechanism of blazars. In the following, we will study the variation of PD,  and the rotation of Stokes parameters in the $qu$ plane in details.
In the four bottom panels of Figure \ref{PDFV}, we show the segmented light curves of {\it V}-band optical flux and PD. It is obvious that the peak times of the flux flares and the PD flares are not synchronous. In the segment (a), the PD firstly rises and then declines during the increasing phase of the flux. We name the PD flare which appears in the rising phase of the flux as the former PD flare. For this segment, the decreasing phase of the flux is not monitored.  There are two flux flares in the segment (b). The first one has only the decreasing phase, accompanied by a complete PD flare, which is named as the latter PD flare. For the second one, the PD peak appears before the flux, which can be classified as the former PD flare.  In the segment (c), the flux first rises, meanwhile the PD decreases. Then the flux declines, and the PD firstly rises and then declines, which is ascribed to be a latter PD flare.  In the segment (d), as the flux decreases, the PD first rises and then declines, which can be identified as the latter PD flare.  In summary, the monotonic variation of the flux will correspond to one complete PD flare. If we assume that one flux flare is sandwiched between the former and latter PD flares, the variations of flux and PD can be understood in a unified manner. However, this assumption is still not qualified from the current data.  Only in the segment (b), one and a half flux flares are monitored, and the PD flares vary more quickly than the flux. \citet{Sasada:2011} showed two complete flux flares during the same period for this target, and more non-simultaneous PD flares indeed were monitored (see Fig. 1 in this literature). There is a caveat that no one complete sandwiched pattern of the PD flares is confirmed from our data. The samplings and the time coverage of the current data are not good enough. Other correspondences between the PD flares and the flux flares are possible.

Two models can explain the observed variation of PD and flux for AO 0235+164. \citet{Nalewajko:2010} presented the polarization produced by the helical motion of the emitting blobs in the jet  model. When the blob swings in our line of sight, the viewing angle $\xi$ between the velocity direction and line of sight will decrease firstly. When the $\xi$ reaches $1/\Gamma$, PD will reach the peak. When $\xi$ approaches the minimum, the optical flux reaches the peak and the PD value decreases. When the blob continues to move along the helical trajectory, $\xi$ will increase and pass through $1/\Gamma$ again, leading to another peak in the PD light curve. Therefore, if the line of sight is inside  the jet cone, an optical flux flare will be sandwiched between the former and latter PD flares.  Another model that can produce such a sandwiched pattern of time series was proposed by \citet{Zhang:2014,Zhang:2016}. This model considers the synchrotron radiation of the proton, and studies the polarization and radiation signatures. The shock moves along the jet and sweeps the helical magnetic field. The observed flux for one sampling is composed of radiation from different regions in the jet, the averaged Stokes parameter will manifest the effective direction of magnetic fields. The enhanced particle injection without changing the topology of the magnetic field will produce two PD flares and one flux flare. The polarization angle (PA) rotation of the first PD flare is the inverse of the second one.

\begin{figure}[ht]
%   \begin{tabular}
        \centerline{\includegraphics[scale=0.8]{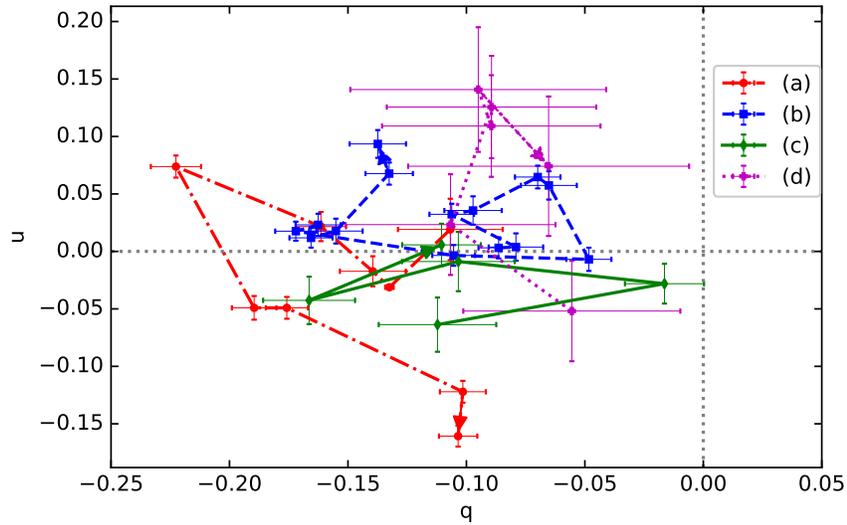}}
         \caption{Behaviors of $q$ and $u$ for the 4 segments are manifested. The arrows at the end of lines show the direction of  $q, u$ rotation.} \label{QU}
\end{figure}

Rather than studying the PA rotation, we study the variation behavior of $(q,u)$, since the PA rotation has the $n-\pi$ ambiguity problem \citep{Marscher:2008,Kiehlmann:2016}, and $q$ and $u$ provide more information than PA. In Figure \ref{QU}, the trajectories of $(q,u)$ corresponding to the four segments are plotted. In the segment (a), the rotation is mainly counterclockwise, except that the first two and last data points show a reflected trend.  In the segment (b), the rotation is first clockwise, and then is changed to counterclockwise. The lines of segments (c) and (d) line mainly show clockwise rotations.  From the analysis above, the likely conclusion is that the former and latter PD flare corresponds to the counterclockwise and clockwise rotation, respectively. There is a caveat that other rotation trends are possible due to the large uncertainty of $q$ and $u$. Such kind of rotations is not likely caused by the random walk process.  One also notes that all lines are in the range of $q<0$, and pass the $u=0$ line. The deviation from the origin of $qu$ plane indicates that another constant polarization component exists. \citet{Sasada:2011}  also presented $qu$ rotations for this target in Epoch I (See Fig.3 in this reference). Their monitoring shows that both the counterclockwise and clockwise rotation appear during the period of flux flares, which agree with our results.
For 3C 454.3, \citet{Sasada:2010} found that a counterclockwise rotation in the $qu$ plane appears for the outburst in 2017, and the clockwise rotations appear in the active state.  \citet{Uemura:2017} studied the  PKS 1749+096, and found the lag between PD and flux light curves. The rotations of ($q,u$) were also observed during many flares. The moving shocks passing through the curved trajectories as well as the Doppler change were proposed to explain the variation phenomena of polarization.

We will present the variations of polarization by using the helical jet model, which was proposed by \citet{Steffen:1995} and investigated by \citet{Mohan:2015,Shablovinskaya:2019}.
The helical jet model respects the conservations of the momentum along the jet axis, the angular momentum, and the kinetic energy. The trajectory of the emitting blob is described by the cylindrical coordinates of $(\rho, \psi, z)$ in the jet coordinate system, where $\rho$ is the distance from the blob to the jet axis, $\phi$ is the azimuth angle, $z$ is the distance from the blob to the jet base. Suppose that the emitting blob is launched at a cylindrical distance $\rho_0$ from the jet base with an initial velocity $\beta=v/c$. The jet half opening angle $\varphi$ is assumed to be constant. The coordinates and the velocities of the emitting blob are given as
\begin{align}
\rho&=f \sqrt{1+\left(\frac{a t+b}{f}\right)^{2}},    \\
\dot{\rho}&=\frac{a}{\rho}(a t+b), \\
\phi&=\left[\arctan \left(\frac{a t+b}{f}\right)-\arctan \left(\frac{b}{f}\right)\right] / \sin \varphi,   \\
\dot{\phi}&=\frac{a f}{\rho^{2} \sin \varphi}, \\
z&=\frac{\rho-\rho_{0}}{\tan \varphi}, \\
\dot{z}&=\frac{\dot{\rho}}{\tan \varphi},
\end{align}
where the constants $a$, $b$, and $f$ are given as $a =\beta \sin \varphi$, $b= \sqrt{\rho_{0}^{2}-f^{2}}$, and $f=j/\beta$ ($j$ is the normalized angular momentum in units of distance).
In the observer system, the angle $\xi$ between the instantaneous velocity direction  and the line of sight is described by \citep{Li:2018}
\begin{equation}
\cos \xi=\frac{\dot{\rho} \cos \phi \sin i-\rho \dot{\phi} \sin \phi \sin i+\dot{z} \cos i}{\left(\dot{\rho}^{2}+\rho^{2} \dot{\phi}^{2}+\dot{z}^{2}\right)^{1 / 2}},
\end{equation}
where $i$ is the angle between the jet axis and the line of sight.
The Doppler factor is $\delta=1/[\Gamma(1-\beta \cos \xi)]$, which changes with time because the angle $\xi$ is time dependent.
\citet{Cawthorne:1990} stated that $\Pi$ (PD) depends on $\xi$, which can be described by the empirical relation $\Pi\sim A\sin^n \xi'$, where $n$ is a positive real number \citep{Shao:2019}. $\Pi$ reaches the maximum for  $\xi'=\pi /2$, and decreases to the minimum for $\xi'= 0$. \citet{Raiteri:2013} suggested that $n=2$. However, we have found that the numerical integrations of Stokes parameters (presented in \citealt{Lyutikov:2005}) are best fitted with the empirical relation for $n=3$. i.e., $\Pi \sim \Pi_{\max } \sin ^{3} \xi^{\prime}$. $\xi'$ can be obtained by the Lorentz transformation from $\xi$ via $\sin \xi^{\prime}=\sin \xi/\Gamma(1-\beta \cos \xi)$.
In the observer system, the electric vector position angle (EVPA) is given as
\begin{equation}
\tan \chi=\frac{e_{u}}{e_{w}}=\frac{\cos \phi \cos i-\tan \varphi \sin i}{\sin \phi},
\end{equation}
where $e_{u}$ and $e_{w}$ are projected values onto the $(u,w)$ reference frame in our sky.  The fractional Stokes parameters $q =Q/I$ and $u = U/I$ can be are expressed as $q=\Pi \cos 2 \chi$ and $u=\Pi \sin 2 \chi$, respectively.

\begin{figure}[ht]
%   \begin{tabular}
    \begin{minipage}[t]{0.48\linewidth}
        \centerline{\includegraphics[scale=0.56]{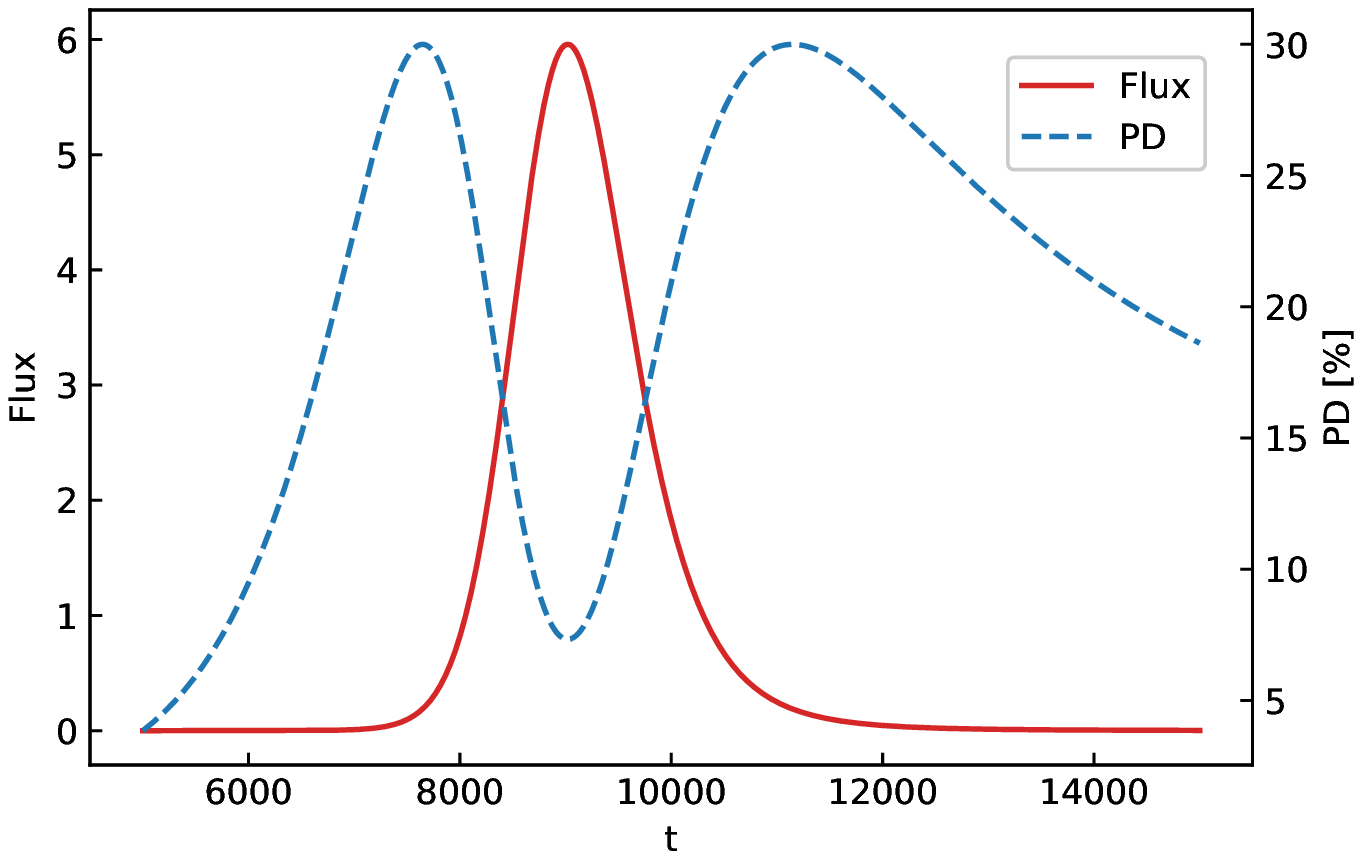}}
           \end{minipage}%
    \begin{minipage}[t]{0.48\linewidth}
        \centerline{\includegraphics[scale=0.56]{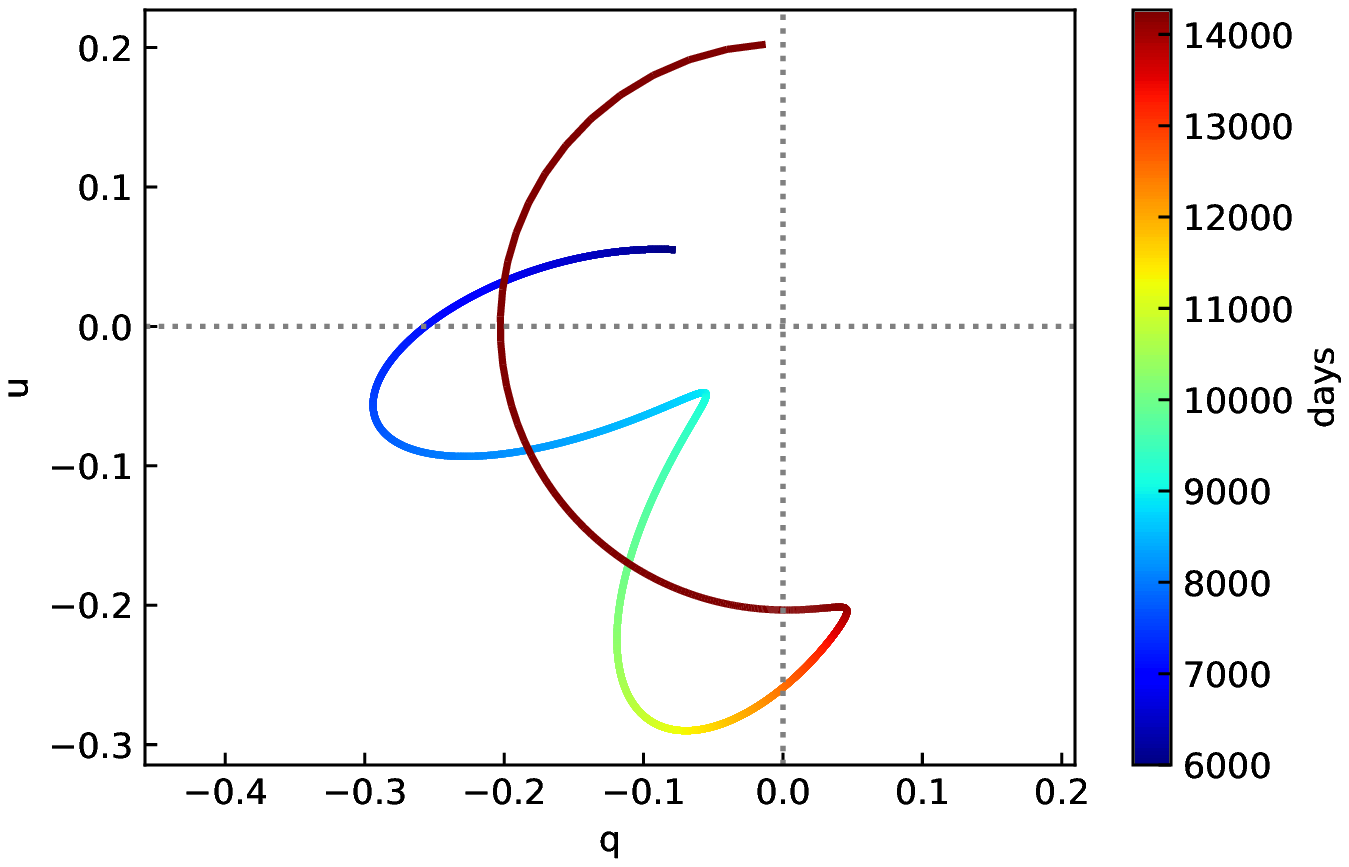}}
    \end{minipage}
\caption{Variation of the flux density, PD, and $(q,u)$ in helical jet model. Left panel: light curves of the flux density (red solid line) and PD (blue dash line). Right panel: rotations of $(q,u)$, the colors correspond to time. Parameters are: Lorentz factor $\Gamma= 14$, inclination angle $i=6^{\circ}$, opening angle $\varphi=2^{\circ}$, normalized angular momentum $j=0.04$, and initial radial distance $\rho_0=7 \times 10^{-2}$ pc. The color of the line in the right panel indicates the time.} %$\Delta t \sim 8$ days  is estimated based on the minimum optical time scale of flux and polarization.}
\label{Fig:model}
\end{figure}

In Figure \ref{Fig:model}, we schematically present the variations of PD, fluxes, and $(q,u)$ using the analytical equations in helical jet model. The input parameters are set as $i=6^{\circ}$, $\varphi=2^{\circ}$, $\Gamma=14$, $j=0.04$, and $\rho_0=0.07$ pc. The parameter $\rho_0$ controls the scale of time. The set of $\rho_0=0.07$ pc leads that the flux flare appears at $t\approx 9000$ days, which roughly agrees with that the optical emitting regions is 8 pc away from the jet base.  Since the viewing angle is larger than the jet opening angle, our line of sight is beyond the jet cone. From the left panel of \ref{Fig:model}, it is evident that the two PD flares sandwich the flux flare as expected. We also check the angle $\xi$, which takes a minimal value $\sim 1.4^{\circ}$ for $t \approx 9000$ days. To study the trajectory of $(q,u)$, we add a constant vector component to $e_u$ and $e_w$, i.e., $e_u+0.8$ and $e_w+0.6$. The idea is that there should be a disk component to explain the RWB trend. And such component can also affect the trajectories of $(q,u)$.
The values of added components are set to realize that $u$ changes sign while $q$ remains negative, as illustrated in Figure \ref{QU}. In the right panel of Figure \ref{Fig:model}, the trajectories of $(q,u)$ shows a counterclockwise rotation for the first PD flare, reflects between two PD flares, shows another counterclockwise rotation for the second flare, and finally exhibits a clockwise rotation at the end of the second PD flare.  \citet{Li:2018} studied the polarization of CTA 102 using the same helical jet model, and obtained the similar PD flares and $qu$ rotations without flipping the rotation trends. We found that the reflection of the rotation trend appears when  $\tan \chi$ passes the singular point, which means that the PA rotation is also flipped. The shock in jet model with a helical magnetic field can also produce the PA variation with sine profile \citet{Zhang:2016}, which may also explain the $qu$ rotations.
Manipulating with other sets of parameters, we obtained that no flip of trends appears, as illustrated by \citet{Li:2018} in the study of CTA 102. Thus, the helical jet model can explain the flip of rotation trend with a fine-tuning of parameters.

Based on the above schematic illustrations, we conclude that observed characteristics of polarization for AO 0235+164 can be explained roughly by the helical jet model. Combining with the analysis of color and the spectral index of $\gamma$-ray, the change of the viewing angle is the primary variable to explain the observed variation phenomena of AO 0235+164. Shock in jet model with helical magnetic filed is an alternative. Better sampling of optical observation, especially the monitoring of polarization,  is indispensable to diagnose models to reveal the variation mechanism.

\section{Conclusion} \label{Sec:conclusion}

In this work, we collected nine years light curves of the $\gamma$-ray, optical {\it V}-band,  radio 15 GHz and polarization. We made the correlation analysis to study the  optical and $\gamma$-ray emitting regions.
We analyze the color and spectral index behavior to investigate the variation mechanism.  The behaviors  of PD and $qu$ rotations enable us to further constrain the models.
The principal conclusions are given as follows

\begin{itemize}
  \item Based on the LCCF analysis, the optical {\it V}-band and $\gamma$-ray light curves are correlated with the radio 15 GHz  with $3\sigma$ significance. The $\gamma$-ray and optical emitting regions are roughly the same, and locate at {$6.6_{-1.7}^{+0.6}$} pc upstream of the core region of 15 GHz. The estimation of the radio core region predicts that the optical and $\gamma$-ray emitting regions are far away from the BLR.

  \item For the variation of color, the target shows an RWB trend at the low flux state, and a BWB trend at the high flux state. While, the spectral index of $\gamma$-ray always shows the SWB trend. With contamination of accretion disk, the increase of the dominant jet component will turn the RWB trend to the BWB trend in a unified manner. The SWB trend for $\gamma$-ray can be explained naturally by the existence  of constant higher energy GeV $\gamma$-ray component. EC process with seed photons from torus is favored to explain the correlation between optical and $\gamma$-ray fluxes. The broad emission lines are evident in the low flux states, which indicates that AO 0235+164 is an FSRQ source.

  \item As a whole, the PDs are not correlated with the optical fluxes. This can be explained by the superposition of multiple flares and the weak disc component. It was found that the PD flares and flux flares are not synchronous. It seems that the counterclockwise and clockwise rotation trend in $qu$ plane corresponds to the former and latter PD flares, respectively. The possibility that such correspondence is an illusion due to the uneven samplings and bad time coverage can not be ruled out. Using the helical jet model, the variations of the flux, PD, and Stokes parameters $(q,u)$ for the target can be simulated schematically.
\end{itemize}

To sum up, AO 0235+164 is an FSRQ target, and the change of the viewing angle can explain its various  phenomena of variation.
\acknowledgments
We are grateful to the anonymous referee for valuable comments which improve the quality of this work.
This work has been funded by the National Natural Science Foundation of China under Grant No. U1531105 and 11403015.
Data from the Steward Observatory spectropolarimetric monitoring project were used. This program is supported by Fermi Guest Investigator grants NNX08AW56G, NNX09AU10G, NNX12AO93G, and NNX15AU81G.
This paper has made use of up-to-date SMARTS optical/near-infrared light curves that are available at www.astro.yale.edu/smarts/glast/home.php
This research has made use of data from the OVRO 40-m monitoring program (Richards, J. L. et al. 2011, ApJS, 194, 29) which is supported in part by NASA grants NNX08AW31G, NNX11A043G, and NNX14AQ89G and NSF grants AST-0808050 and AST-1109911.

\vspace{5mm}
\facilities{Fermi(LAT), Steward, SMARTS, OVRO:40m}

%
%\begin{figure}[ht]
%%   \begin{tabular}
%    \begin{minipage}[t]{0.48\linewidth}
%        \centerline{\includegraphics[scale=0.5]{chit.eps}}
%           \end{minipage}%
%    \begin{minipage}[t]{0.48\linewidth}
%        \centerline{\includegraphics[scale=0.5]{xit.eps}}
%    \end{minipage}
%\caption{Variation of the flux density, PD, and $(q,u)$ in helical jet model. Left panel: light curves of the flux density (blue line) and PD (red line). Right panel: rotations of $(q,u)$. Parameters are: Lorentz factor $\Gamma= 14$, inclination angle $i=6^{\circ}$, opening angle $\varphi=1.6^{\circ}$, normalized angular momentum $j=0.004$, and initial radial distance $\rho_0=7 \times 10^{-2}$ pc.} %$\Delta t \sim 8$ days  is estimated based on the minimum optical time scale of flux and polarization.}
%\end{figure}


\begin{thebibliography}{}

\bibitem[Abdo et al.(2009)]{Abdo:2009}Abdo, A. A., Ackermann, M., Ajello, M., et al. 2009, \apjs, 183, 46

\bibitem[Acero et al.(2015)]{Acero:2015}Acero, F., Ackermann, M., Ajello, M., et al. 2015, \apjs, 218, 23

\bibitem[Ackermann et al.(2012)]{Ackermann:2012}Ackermann, M., Ajello, M., Ballet, J., et al. 2012, \apj, 751, 159

\bibitem[Agudo et al.(2011)]{Agudo:2011}Agudo, I., Marscher, A., Jorstad, S., et al. 2011, \apjl, 735, L10

\bibitem[Agudo(2013)]{Agudo:2013}Agudo, I. 2013, EPJWC, 61, 04002

\bibitem[Atwood et al.(2009)]{Atwood:2009} Atwood, W. B., Abdo, A. A., Ackermann, M., et al. 2009, \apj, 697, 1071

\bibitem[Bonning et al.(2012)]{Bonning:2012}Bonning, E., Urry, C. M., Bailyn, C., et al. 2012, \apj,  756, 13

\bibitem[B\"{o}ttcher et al.(2013)]{Bottcher:2013} B\"{o}ttcher, M., Reimer, A., Sweeney, K., \& Prakash, A. 2013, \apj, 768, 54

\bibitem[Cawthorne \& Cobb(1990)]{Cawthorne:1990}Cawthorne, T. V., \& Cobb, W. K. 1990, \apj, 350, 536

\bibitem[Cellone et al.(2007)]{Cellone:2007}Cellone, S. A., Romero, G. E., Combi, J. A., Mart\'{\i}, J. 2007, \mnras, 381, L60

\bibitem[Celotti \& Ghisellini(2008)]{Celotti:2008}Celotti, A., \& Ghisellini, G. 2008, \mnras, 385, 283

\bibitem[Chen \& Jiang(2001)]{Chen:2001}Chen, Y. J., \& Jiang, D. R. 2001, \aap, 376, 69

\bibitem[Chen et al.(2002)]{Chen:2002}Chen, Y. J., Jiang, D. R. , \& Zhang, F. J. 2002, \aap, 26, 14

\bibitem[Ciprini(2015)]{Ciprini:2015}Ciprini, S. 2015, ATel, 7975

\bibitem[Cohen et al.(1987)]{Cohen:1987}Cohen, R. D., Smith, H. E., Junkkarinen, V. T., Burbidge, E. M. 1987, \apj, 318, 577

\bibitem[Corbel et al.(2008)]{Corbel:2008}Corbel, S., \& Reyes, L. C. 2008, ATel, 1744

\bibitem[Davis et al.(1982)]{Davis:1982}Davis, M. M., Wolfe, A. M. 1982, IAUS, 97, 311

\bibitem[Edelson \& Krolick(1988)]{Edelson:1988}Edelson, R. A., \& Krolik, J. H. 1988, \apj, 333, 646

\bibitem[Elvis et al.(1994)]{Elvis:1994}Elvis, M., Wilkes, B. J., McDowell, J. C., et al. 1994, \apjs, 95, 1

\bibitem[Fan et al.(2017)]{Fan:2017}Fan, J. H., Kurtanidze, O., Liu, Y., et al. 2017, \apj, 837, 45

\bibitem[Foschini et al.(2008)]{Foschini:2008}Foschini, L., Longo, F., \& Iafrate, G. 2008, ATel, 1784

\bibitem[Hagen-Thorn et al.(2008)]{Hagen:2008}Hagen-Thorn, V. A., Larionov, V. M., Jorstad, S. G., et al. 2008, \apj, 672, 40

\bibitem[Hagen-Thorn et al.(2018)]{Hagen:2018}Hagen-Thorn, V. A., Larionov, V. M., Morozova, D. A., et al. 2018, ARep, 62, 103

\bibitem[Hirotani(2005)]{Hirotani:2005}Hirotani, K. 2005, \apj, 619, 73

\bibitem[Hovatta et al.(2009)]{Hovatta:2009}Hovatta, T., Valtaoja E., Tornikoski, M., L\"{a}hteenm\"{a}ki, A. 2009, \aap, 494,527

\bibitem[Ikejiri et al.(2011)]{Ikejiri:2011}Ikejiri, Y., Uemura, M., Sasada, M., et al. 2011, \pasj, 63, 639

\bibitem[Jenkins(1996)]{Jenkins:1996}Jenkins, P. 1996, ASPC, 110, 156

\bibitem[Jiang et al.(2016)]{Jiang:2016}Jiang, Y. G., Hu, S. M., Chen, X., et al. 2016, \mnras, 456, 3386

\bibitem[Jiang et al.(2020)]{Jiang:2020}Jiang, Y. G., Hu, S. M., Chen, X., Shao, X., \& Huo, Q. H.  2020, \mnras, 493, 3757

\bibitem[Johnston et al.(1995)]{Johnston:1995}Johnston, K. J., Fey, A. L., Zacharias, N., et al. 1995, AJ, 110, 880

\bibitem[Jorstad et al.(2017)]{Jorstad:2017}Jorstad, S. G., Marscher, A. P., Morozova, D. A., et al. 2017, \apj, 846, 98

\bibitem[Kiehlmann et al.(2016)]{Kiehlmann:2016}Kiehlmann, S., Savolainen, T., Jorstad, S. G. et al. 2016, \aap, 590, A10

\bibitem[Kirk et al.(1998)]{Kirk:1998}Kirk, J. G., Rieger, F. M., \& Mastichiadis, A. 1998, \aap, 333, 452

\bibitem[K\"{o}nigl(1981)]{Konigl:1981}K\"{o}nigl, A. 1981, \apj,  243, 700

\bibitem[Kraus et al.(1999)]{Kraus:1999}Kraus, A., Quirrenbach, A., Lobanov, A. P., et al. 1999, \aap, 344, 807

\bibitem[Kudryavtseva et al.(2011)]{Kudryavtseva:2011}Kudryavtseva, N. A., Gabuzda, D.C., Aller, M. F., \& Aller, H.D. 2011, \mnras, 415, 1631

\bibitem[Kutkin et al.(2018)]{Kutkin:2018} Kutkin, A. M., Pashchenko, I. N., Lisakov, M. M., et al. 2018, \mnras, 475, 4994

\bibitem[Larsson(2012)]{Larsson:2012}Larsson, S., 2012, arxiv:1207.1459v1

\bibitem[Ledden et al.(1976)]{Ledden:1976}Ledden, J. E., Aller, H. D., \& Dent, W. A. 1976,  {Nature}, 260, 752

\bibitem[Li et al.(2018)]{Li:2018}Li, X. F., Mohan, P., An, T., et al. 2018, \apj, 854, 17

\bibitem[Lister et al.(2009)]{Lister:2009}Lister, M. L., Aller, H. D., Aller, M. F., et al. 2009, \aj, 137, 3718

\bibitem[Liu et al.(2006)]{Liu:2006}Liu, F. K., Zhao, G., Wu, X. B. 2006, \apj, 650, 749

\bibitem[Lobanov(1998)]{Lobanov:1998}Lobanov, A. P. 1998, \aap,  330, 79

\bibitem[Lyutikov et al.(2005)]{Lyutikov:2005}Lyutikov, M., Pariev, V. I., \& Gabuzda, D.C. 2005, \mnras, 360, 869

\bibitem[Macleod et al.(1976)]{Macleod:1976}MacLeod, J. M., Andrew, B. H., \& Harvey, G. A. 1976,  {Nature}, 260, 751

\bibitem[Marscher et al.(2008)]{Marscher:2008}Marscher, A. P., Jorstad, S. G.,  D'Arcangelo, F. D., et al. 2008,  {Nature}, 452, 966

\bibitem[Max-Moerbeck et al.(2014a)]{Max:2014a}Max-Moerbeck, W., Hovatta, T., Richards, J. L., et al. 2014a, \mnras, 445, 428

\bibitem[Max-Moerbeck et al.(2014b)]{Max:2014b}Max-Moerbeck, W., Richards, J. L., Hovatta, T., et al. 2014b, \mnras, 445, 437

\bibitem[Mohan et al.(2015)]{Mohan:2015}Mohan, P., Agarwal, A., Mangalam, A., et al. 2015, \mnras, 452, 2004

\bibitem[M\"{u}cke et al.(2003)]{Mucker:2003}M\"{u}cke, A., Protheroe, R. J., Engel, R., Rachen, J. P., \& Stanev, T. 2003, APh, 18, 593

\bibitem[Nalewajko(2010)]{Nalewajko:2010}Nalewajko, K. 2010, IJMPD, 19, 701

\bibitem[Ostorero et al.(2004)]{Ostorero:2004}Ostorero, L., Villata, M., \& Raiteri, C. M. 2004, \aap, 419, 913

\bibitem[O'Sullivan \& Gabuzda(2009)]{OSullivan:2009}O'Sullivan, S. P., \& Gabuzda, D. C. 2009, \mnras, 400, 26

\bibitem[Peterson et al.(1998)]{Peterson:1998}Peterson, B. M., Wanders, I., Horne, K., et al. 1998, \pasp, 110, 660

%\bibitem[Pushkarev et al.(2012)]{Pushkarev:2012}Pushkarev, A. B.,  {Hovatta, T., Kovalev, Y. Y.,} et al. 2012, \aap, 545,  {A113}

\bibitem[Rain\`{o} et al.(2013)]{Raino:2013}Rain\`{o}, S., Madejski, G., do Couto e Silva, E., et al. 2013, NuPhS, 239, 270

\bibitem[Raiteri et al.(2001)]{Raiteri:2001}Raiteri, C. M., Villata, M., Aller, H. D., et al. 2001, \aap, 377, 396

\bibitem[Raiteri et al.(2006)]{Raiteri:2006}Raiteri, C. M., Villata, M., Kadler, M., et al. 2006, \aap, 459, 731

\bibitem[Raiteri et al.(2007)]{Raiteri:2007}Raiteri, C. M., Villata, M., Capetti, A., et al. 2007, \aap, 464, 871


\bibitem[Raiteri et al.(2013)]{Raiteri:2013}Raiteri, C. M., Villata, M., D'Ammando, F., et al. 2013, \mnras, 436, 1530

\bibitem[Raiteri et al.(2017)]{Raiteri:2017}Raiteri, C. M., Villata, M., Acosta-Pulido, J. A., et al. 2017,  {Nature}, 552, 374

\bibitem[Richards et al.(2011)]{Richard:2011}Richards, J. L. Max-Moerbeck, W., Pavlidou, V., et al. 2011, \apjs, 194, 29

\bibitem[Rieke et al.(1976)]{Rieke:1976}Rieke, G. H., Grasdalen, G. L., Kinman, T. D., et al. 1976,  {Nature}, 260, 754

\bibitem[Sasada et al.(2010)]{Sasada:2010}Sasada, M., Uemura, M., Arai., A., et al. 2010, \pasj, 62, 645

\bibitem[Sasada et al.(2011)]{Sasada:2011}Sasada, M., Uemura, M., Fukazawa, Y., et al. 2011, \pasj, 63, 489

\bibitem[Shablovinskaya \& Afanasiev(2019)]{Shablovinskaya:2019} Shablovinskaya, E., \& Afanasiev, V. 2019, \mnras, 482, 4322

\bibitem[Shao et al.(2019)]{Shao:2019}Shao, X., Jiang, Y. G., \& Chen, X. 2019, \apj, 884, 15

\bibitem[Sikora et al.(1994)]{Sikora:1994}Sikora, M., Begelman, M., \& Rees, M. 1994, \apj, 421, 153

\bibitem[Sikora et al.(2009)]{Sikora:2009}Sikora, M., Stawarz, {\L}., Modersiki, R., Nalewajko, K., \& Madejski, G. 2009, \apj, 704, 38

\bibitem[Smith et al.(2009)]{Smith:2009}Smith, P. S., Montiel, E., Rightley, S., et al. 2009, arXiv:0912.3621

\bibitem[Spinrad \& Smith(1975)]{Spinrad:1975}Spinrad, H., \& Smith, H. E. 1975, \apj, 201, 275

\bibitem[Steffen et al.(1995)]{Steffen:1995}Steffen, W., Zensus, J. A., Krichbaum, T. P., Witzel, A., \& Qian, S. J. 1995, \aap, 302, 335

\bibitem[Stein et al.(1976)]{Stein:1976}Stein, W. A., O'Dell, S. L., \& Strittmatter, P. A. 1976, \araa, 14, 173

\bibitem[Timmer \& K\"{o}nig(1995)]{Timmer:1995}Timmer, J., \& K\"{o}nig, M. 1995, \aap, 300, 707

\bibitem[Uemura et al.(2017)]{Uemura:2017}Uemura, M., Itoh, R., Liodakis, I., et al. 2017, \pasj, 69, 96

\bibitem[Urry \& Padovani(1995)]{Urry:1995}Urry, C. M., \& Padovani, P. 1995, \pasp, 107, 803

\bibitem[Villata\& Raiteri(1999)]{Villata:1999}Villata, M., \& Raiteri, C. M. 1999, \aap, 347, 30

\bibitem[Villata et al.(2006)]{Villata:2006}Villata, M., Raiteri, C. M., Balonek, T. J., et al. 2006, \aap, 453, 817

\bibitem[Villata et al.(2009)]{Villata:2009}Villata, M., Raiteri, C. M., Larionov, V. M., et al. 2009, \aap, 501, 455

\bibitem[Vol'vach et al.(2015)]{Volvach:2015}Vol'vach, A. E., Larionov, M. G., Vol'vach, L. N., et al. 2015, ARep, 59, 145

\bibitem[Wang et al.(2014)]{Wang:2014}Wang, H. T. 2014, \apss, 351, 281

\bibitem[Welsh(1999)]{Welsh:1999}Welsh, W. F. 1999, \pasp, 111, 1347

\bibitem[White \& Peterson(1994)]{White:1994}White, R. J., \& Peterson, B. M. 1994, \pasp, 106, 879

\bibitem[Zhang et al.(2014)]{Zhang:2014}Zhang, H. C., Chen, X., \& B\"{o}chtter, M. 2014, \apj, 789, 66

\bibitem[Zhang et al.(2016)]{Zhang:2016}Zhang, H. C., Diltz, C., \& B\"{o}chtter, M. 2016, \apj, 829, 69


\end{thebibliography}
\end{document}